\documentclass[journal]{IEEEtran}
\usepackage{algorithmic}
\usepackage{amsmath}
\usepackage{amsfonts}
\usepackage[pdftex]{graphicx}
\graphicspath{{fig/}}
\usepackage{threeparttable}
\usepackage{booktabs}
\usepackage[linesnumbered, ruled, vlined]{algorithm2e}
\usepackage{cite}
\usepackage[hyphens]{url}
\usepackage{hyperref}
\usepackage{makecell}
\usepackage{xcolor}

\begin{document}

\title{Recent Advances in Resilient Multi-Energy Systems Against Climate Change
\thanks{G. Ruan is with the Laboratory for Information \& Decision Systems, Massachusetts Institute of Technology, Boston, MA 02139, USA. He was previously with The University of Hong Kong. (email: gruan@mit.edu, orcid: 0000-0003-2660-9298)}
\thanks{Z. Li is with the School of Electrical Engineering, Aalto University, Espoo 02150, Finland. (email: zhengmao.li@aalto.fi, orcid: 0009-0008-1851-9800)}
\thanks{Y. Wang is with the Department of Electrical and Computer Engineering, The University of Hong Kong, Hong Kong SAR, China. (email: yiwang@eee.hku.hk, orcid: 0000-0003-1143-0666)}
\thanks{N. Zhang is with the Department of Electrical Engineering, Tsinghua University, Beijing 100084, China. (email: ningzhang@tsinghua.edu.cn, orcid: 0000-0003-0366-4657)}
}

\author{
Grant~Ruan, \quad
Zhengmao~Li, \quad
Yi~Wang, \quad
Ning~Zhang \quad~~~~
}

\markboth{Accepted by Proceedings of the IEEE}{}

\maketitle

\begin{abstract}
Climate change is a global threat to the long-term sustainable development of energy systems. Recent works have explored the emerging opportunity of coordinating different energy carriers and sectors (e.g. electricity, natural gas, heating, hydrogen, transportation, and water sectors) to unlock the cross-sector flexibility against climate change. 
This review has established a holistic framework for resilient multi-energy systems through the lens of nested coupling. It covers the most recent progress in resilience resources, resilience evaluation, resilience-oriented operation \& planning, resilience pricing \& investment, and real-world implementation. This work differs from prior studies through a full investigation on climate change impacts (distribution shifts), resilience pricing, and global projects. Within this area, we advocate for a unique and interdisciplinary perspective spanning across energy systems, climate science, sociology, economics, and data science. At the end, seven major challenges and opportunities are identified, including data deficiency, distributed coordination \& privacy, high-fidelity simulation, and machine learning techniques. Researchers, industrial experts, and policy makers can follow this review to capture the emerging trend and future opportunities in this growing area.
\end{abstract} 

\begin{IEEEkeywords} 
resilience, reliability, integrated energy systems, energy hubs, extreme/adverse weather, risk management, renewable energy, microgrid, infrastructure
\end{IEEEkeywords}

\section{Introduction} \label{sec:intro}

\subsection{Background and Motivation}

Climate change has emerged as a significant global crisis that causes extreme weather events, food shortages, and species extinction. The Intergovernmental Panel on Climate Change (IPCC) reported that around 3.3--3.6 billion people lived in places highly vulnerable to climate change~\cite{core2024synthesis}. Climate change is even progressing faster than past expectations, with the National Oceanic and Atmospheric Administration~(NOAA) warning that the ten warmest years on record (global temperature) occurred during 2015–2024~\cite{noaa2024annual}. 

The world has to face growing dangers from record-breaking weather events~\cite{ritchie2025ourworld}, including the Southern California wildfires~(2025), the European heat wave~(2024), and the Rio Grande do Sul floods~(2024). According to the global Emergency Event Database, 393 extreme events from 2024 could account for 16,753 deaths, 167.2 million victims, and a total economic loss of \$241.95 billion~\cite{emdat2024disasters}. The World Meteorological Organization released a 50-year assessment report to share that the economic loss rose 48.3\% from \$997.9 billion in 2000--2009 to \$1.48 trillion in 2010--2019~\cite{wmo2023atlas}. 

At the meanwhile, modern energy infrastructures are undergoing a major landscape shift towards deep decarbonization. A growing number of fossil fuel generators are overwhelmingly replaced by less-controllable and weather-driven renewable energy generators~\cite{ghanbarzadeh2025addressing}. According to the International Energy Agency (IEA), renewable penetration across the globe was projected to reach 2.6 times by 2030 (2022 baseline)~\cite{iea2025renewables}. Such a huge change in the generation portfolio may hinder risk management against extreme weather and increase energy-system vulnerability. In recent years, this issue has already become one of the top challenges.

Motivated by the above facts, a novel resilience framework is required to understand the energy system vulnerability under climate change. This framework should align with the sector coupling and demonstrate how disturbances may transit across sectors and cause cascading failures.

\subsection{Real-World Events} \label{subsec:real-world-events}

\begin{figure}
	\centering
	\includegraphics[width=0.31\textwidth]{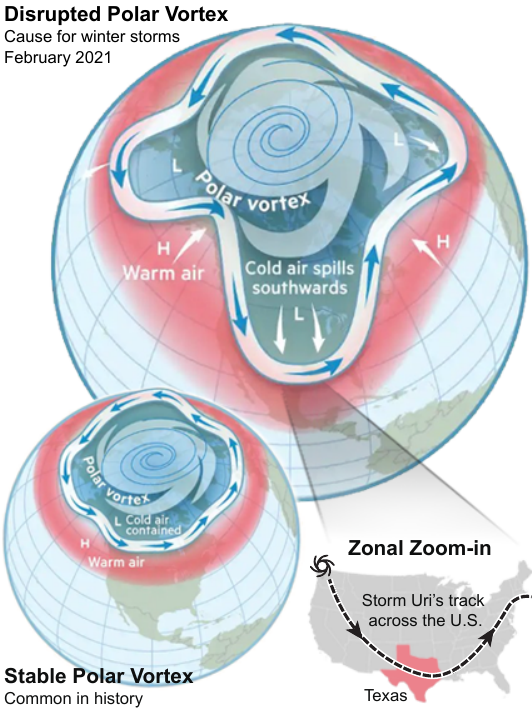}
	\caption{Polar vortexes and the track of winter storm Uri across the mainland. In an unusual and disrupted polar vortex, the jet stream becomes wavy and cool polar air can blow southward to cause extreme cold weather in mid-latitude regions. This figure is inspired by~\cite{hook2021how}.}
	\label{fig:vertox}
\end{figure}

Multiple energy system failures occurred in recent years. The Texas blackout crisis in February 2021 must be among the best-known incidents. Fig.~\ref{fig:vertox} explains how a disrupted polar vortex can cause winter storms in Texas~\cite{hook2021how}.

The Public Utility Commission of Texas (PUCT)~\cite{utaus2021timeline} reported at least 57 deaths across the state, 4.5 million customers without power, and \$195 billion of infrastructure damage. It was observed that the maximal load shedding during the blackout was 20~GW, which is 26\% of the regular peak load. Although Texas had abundant wind energy (24.8\% of the entire generation resources~\cite{wu2021open}), wind generators suffered a few early outages during the storms due to ice coverage on blades, gearboxes, and nacelles. An 8--20~GW outage of wind power was attributed to be the trigger and a primary cause of the blackout. Another important finding from~\cite{utaus2021timeline} was that the energy system dependence further intensified the blackout. In particular, the natural gas production, distribution, and storage infrastructures failed to provide the full amount of natural gas to run gas-fired power plants (85\% drop in dry gas production) before mid-February. This extreme event in Texas has shown an urgent need to integrate resilience thinking into energy infrastructure designs, especially as renewables and energy system dependence are still in steady growth.

The Texas blackout crisis (2021) is unfortunately not an isolated case. Below, we select a group of infrastructure failures with a duration ranging from days to months:

i)~Tohoku Earthquake and Tsunami~(2011)~\cite{aki2017demand}: In March, the earthquake (9.1 magnitudes) and tsunami killed 19 thousand people in Japan and interrupted the electricity supply for 8.7 million residents, natural gas for 0.46 million, and water for 2.3 million. After the shutdown of nuclear power stations, rolling blackouts took place in Tokyo for ten days and decreased train services by 30-50\%.

ii)~Hurricane Isaac~(2012)~\cite{eia2012survey}: In late August, the hurricane disrupted over 74\% of the total natural gas processing capacity in Southern Texas. This disruption caused upstream supply shortage and downstream infrastructural damage, which led to widespread electric power outages affecting nearly 0.89 million users in Louisiana.

iii)~Superstorm Sandy~(2012)~\cite{haraguchi2016critical}: In late October, the superstorm and the accompanying floods severely damaged the U.S. northeastern coast with over 200 people killed and 8.5 million out of power. In New York City, risks from the electricity sector propagated to the other interconnected sectors of petroleum, transportation, and communication. Using spatial scope, the total indirect damage due to sector coupling was found greater than the direct damage. 

iv)~Australia Bushfires~(2019)~\cite{civil2020whole}: Since September, the destructive bushfires had hit the Australian east coast for over six months. This is a typical compound disaster with 200 concurrent and sequential fires. Tens of thousands of homes were out of power. These outages and other direct destruction collectively resulted in cascading failures of transportation, communication, and water treatment.

v)~Hawaii Wildfires~(2023)~~\cite{wfca2024after}: In August, four major wind-driven wildfires broke out in Hawaii (Maui island). These massive wildfires ranked among the top ten and caused an estimated property loss of \$6 billion. Poor access roads delayed the utility maintenance vehicles, and water distribution stations were vulnerable with no grid-side power supply.

\begin{figure}
	\centering
	\includegraphics[width=0.42\textwidth]{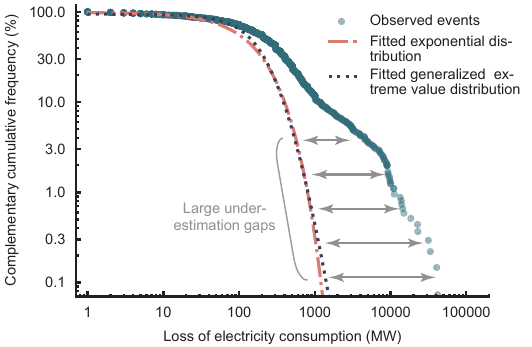}
	\caption{Electricity consumption loss of the U.S. outages during 2000--2023. For each data point ($x$, $y$), $x$ denotes the demand loss and $y$ denotes a complementary cumulative probability that losses are larger than $x$.}
	\label{fig:outage}
\end{figure}

Beyond the above real-world incidents, the destructive impacts of extreme events might also be higher than any past estimations. In Fig.~\ref{fig:outage}, the Form OE-417~\cite{doe2022electric} is collected to study the distributions of energy loss since 2020. We calibrate two heavy-tailed distributions to match the 0--75\% range of data, but the large gaps (double arrows) reflect a significant underestimation of the top 25\% extreme events. 

Clearly, the energy community can no longer rely on rule-based emergency plans for extreme events; instead, we must better understand worst cases and take adaptive actions to mitigate climate-induced risks. Establishing a novel resilience framework that considers climate adaptation and sector coupling is thus an immediate priority.

\subsection{Research Trend and Scope}

\begin{figure}
	\centering
	\includegraphics[width=0.43\textwidth]{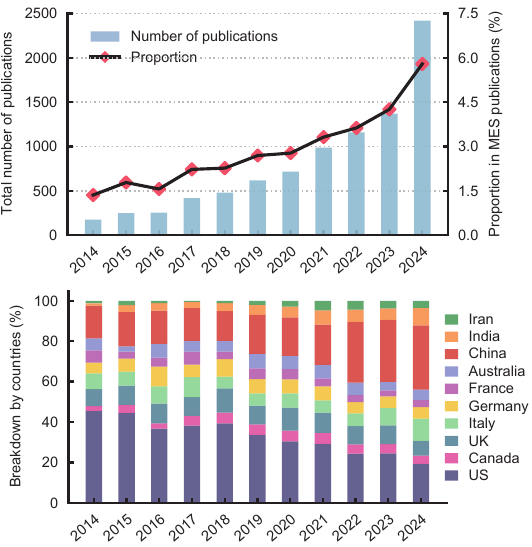}
	\caption{Research trend of multi-energy system resilience in Web of Science database. The top panel shows publication numbers and proportions, with country-level breakdowns below.}
	\label{fig:bib-stats}
\end{figure}

Multi-energy system resilience serves as a versatile framework to capture energy coupling and climate change adaptation. There is a growing research interest in this area over recent years. As indicated in Fig.~\ref{fig:bib-stats}, the publications in Web of Science~(WOS) over the last ten years reflect a linear trendency in the publication number and proportion. It also shows that this area was first led by the U.S., but Asian countries such as China, India, and Iran soon caught up. 

This paper is focused on the resilience of multi-energy systems against climate change for different stakeholders. We focus on how energy dependence and climate change patterns may influence resilience performance, how to model these impacts with high precision, and how to utilize the cross-sector resilience resources to secure the energy supply. To address these challenges, we intend to apply an interdisciplinary perspective regarding engineering, climate science, sociology, economics, and data science.

\begin{figure*}
	\centering
	\includegraphics[width=0.85\textwidth]{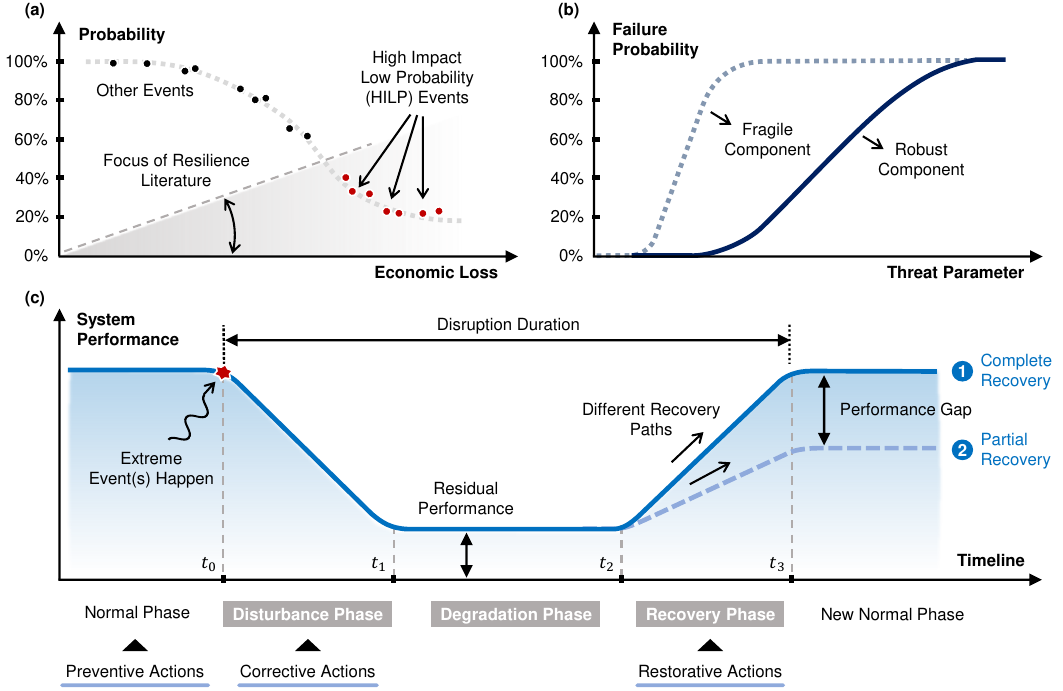}
	\caption{Illustration of the key concepts in resilience literature. These concepts are (a) high impact low probability events, (b) fragility curves, and (c) resilience curves. Detailed annotations are given within each subfigure to provide further demonstrations.}
	\label{fig:key-concepts}
\end{figure*}

The scope includes all the resilience phases/actions and all the sub-categories of operational/structural resilience, short-/long-term resilience, and engineering/economic resilience. We cover natural disasters rather than man-made attacks to keep the main context more concentrated. Fig.~\ref{fig:key-concepts} visualizes three key concepts in the literature: high-impact low-probability (HILP) events, fragility curves, and resilience curves.

\subsection{Contributions}

This paper summarizes the recent advances in multi-energy system resilience that are seldom covered by the existing reviews. Beyond power system resilience~\cite{huang2024toward}, the most relevant reviews~\cite{jasiunas2021energy} and \cite{yang2022resilience} were focused more on conceptual discussions instead of technical modeling. Many efforts lack insights of climate change and resilience investment as well. A thorough comparison is deferred to Appendix~\ref{app:comp-review}.

Overall, our major contributions are threefold:
\begin{enumerate}
\item This paper conducts a comprehensive review on resilient multi-energy systems, covering the potential resources, enhancement strategies, evaluation metrics, operation \& planning models, investment management, and practical implementations. We pay special attention to how energy coupling influence system resilience.

\item An extended investigation on climate change impacts, resilience pricing schemes, and global projects is provided from an interdisciplinary perspective. A special discussion of nested coupling and distribution shifts is made to illustrate climate change impacts. All of these additional efforts enrich the current knowledge. 

\item Emerging challenges and opportunities of data deficiency, distributed coordination, privacy protection, realistic simulations, machine learning applications, and open benchmarks are explored to inspire future works.
\end{enumerate}

\section{Framework} \label{sec:framework}


\subsection{Nested Coupling}

Resilient multi-energy systems requires deep understanding of nested coupling, where diverse energy pathways and temporal/spatial scales are interconnected. Such coupling, arising from energy-source diversification and widespread electrification, is further intensified by climate change.

Two major factors of coupling are considered: sector coupling (internal factor) and coupled disruptions (external factor). The first factor integrates different energy sectors to enhance flexibility and resilience, while the second captures simultaneous or cascading failures across sectors. 

Sector coupling and coupled disruptions can be intertwined in multi-energy systems. Traditionally, energy systems rely on redundant capacities for resilience performance, and sector coupling offers a cost-effective way to share idle redundancy through energy conversion. However, this saving will be limited or invalid if the disruptions are concurrent. Since coupled disruptions are common in many climate-induced accidents (see Subsection~\ref{subsec:real-world-events}), we need to reexamine the interconnected risks and resilience analysis under climate change.


\subsection{A Holistic Framework}

\begin{figure*}
	\centering
	\includegraphics[width=0.83\textwidth]{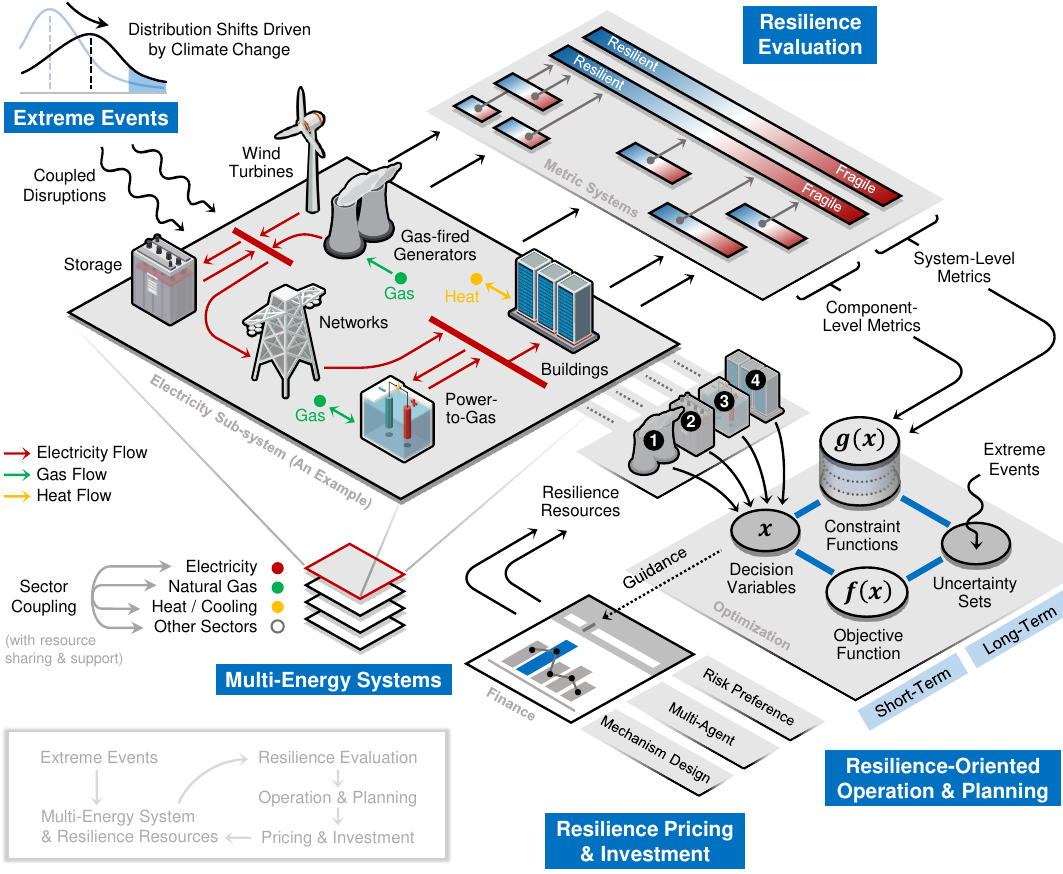}
	\caption{A holistic framework for multi-energy system resilience research. Five major topics include extreme event modeling, multi-energy system \& resilience resources, resilience evaluation, resilience-oriented operation \& planning, and resilience pricing \& investment.} 
	\label{fig:framework}
\end{figure*}

Fig.~\ref{fig:framework} established a holistic framework to study nested coupling. Here, the coupled disruptions are reflected in the extreme event modeling, and the sector coupling is extended and detailed for different resilience tasks. Five major topics are sketched in the lower left corner. 

Within this framework, the extreme event modeling captures the frequency and underlying mechanisms of climate-driven risks. Multi-energy systems apply resilience resources to avoid catastrophic losses, where multi-energy coordination is required for effective use. Resilience evaluation then quantifies performance and supports operational \& planning decisions, as well as resilience pricing \& investment. At last, the investment expands the pool of resilience resources and enables infrastructure retrofits.

Note that this framework extends beyond engineering resilience and comprehensively integrates insights from climate science, data science, and economics.

Below is a navigation for the rest technical sections:
\begin{itemize}
\item \textbf{\ref{sec:extreme-event}. Extreme Events Driven by Climate Change}:
Distribution shifts \ref{subsec:dist-shift}, extreme event descriptions (\ref{subsec:cat-and-des}), extreme event modeling \& prediction (\ref{subsec:mod-and-pred}), and impact analysis on system components (\ref{subsec:imp-sys-comp}).

\item \textbf{\ref{sec:resource}. Multi-Energy Coordination and Resilience Resources}: 
Multi-energy system models (\ref{subsec:evol-mes-model}), potential resilience resources (\ref{subsec:res-resource}), and resilience enhancement strategies (\ref{subsec:res-enh-strategy}).

\item \textbf{\ref{sec:evaluation}. Resilience Evaluation of Multi-Energy Systems}: 
Component-level metrics (\ref{subsec:com-level-metric}), system-level metrics (\ref{subsec:sys-level-metric}), and comprehensive evaluation (\ref{subsec:comp-eval}).

\item \textbf{\ref{sec:optimization}. Resilience-Oriented Operation and Planning of Multi-Energy Systems}: 
Canonical optimization forms (\ref{subsec:canoical-form}), typical solutions (\ref{subsec:res-solution}), resilience-oriented operation (\ref{subsec:res-operation}) and planning (\ref{subsec:res-planning}).

\item \textbf{\ref{sec:investment}. Resilience Pricing and Investment for Multi-Energy Systems}: 
Resilience pricing \& incentive design (\ref{subsec:price-incentive}) and resilience investment (\ref{subsec:res-investment}).

\item \textbf{\ref{sec:project}. Practical Implementation}: 
Real-world projects and implementation in Europe (\ref{subsec:proj-europe}), the U.S. (\ref{subsec:proj-na}), and Asia (\ref{subsec:proj-asia}).

\item \textbf{\ref{sec:challenge}. Emerging Challenges and Opportunities}: 
Data insufficiency of extreme event scenarios (\ref{subsec:dist-shift-ext}), distributed coordinated \& privacy protection (\ref{subsec:dist-coord-privacy}), reliable resilience evaluation (\ref{subsec:rel-res-eval}), realistic simulation (\ref{subsec:real-simul}), investment pricing \& regulation (\ref{subsec:scar-prc-inv}), machine learning applications (\ref{subsec:dl-res-anls}), and open benchmarks (\ref{subsec:benchmark}). 
\end{itemize}

\section{Extreme Events Driven by Climate Change} \label{sec:extreme-event}

This section uses distribution shifts to show climate change impacts. Details of extreme event modeling are followed.

\subsection{Distribution Shifts} \label{subsec:dist-shift}

A fundamental impact of climate change is distribution shift, where meteorological data distributions may change over time and make past experience less predictive for future trends.
This poses a major challenge to the trend extrapolation that suffer from severe bias due to poor generalization. 

Detecting distribution shifts is crucial in extreme weather attribution. A general observation was that the frequency and intensity of natural disasters did not follow a Gaussian distribution but a long-tailed distribution~\cite{arif2021evolving}. Since fewer data points could be captured in the distributional tails, the distribution shift might severely harm the estimation.

\begin{figure}
	\centering
	\includegraphics[width=0.41\textwidth]{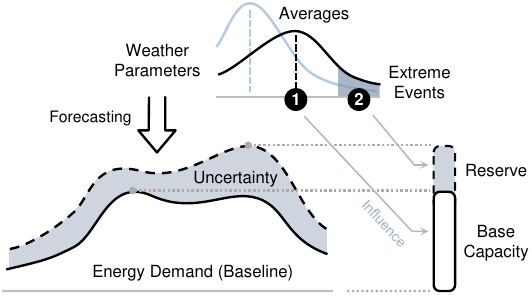}
	\caption{Impacts of distribution shifts. Average drifts shape base capacity, and extreme events determine reserve  needs (redundancy).}
	\label{fig:distrib-shift}
\end{figure}

Distribution shifts are described by two factors: intensified extreme events and drifting mean values. Any high-dimensional meteorological variables can be decomposed into the mean and variation, which are useful to track their changes. For instance, the two factors were recognized as low impact variations and extreme events in \cite{perera2020quantifying}, and average observations and outliers in \cite{levin2022extreme}. Fig.~\ref{fig:distrib-shift} visualizes how they are connected to the system reserve/backup capacity.

\subsubsection{Intensified Extreme Events}

Extreme events may become more severe, more destructive, and more frequent than in the past. A typical phenomena is called compound events~\cite{fehlman2025scoping}, which mean two or more extreme events happen simultaneously or successively. Reference~\cite{ravadanegh2022multi} proposed the concept of multi-harzards disasters. Although historically rare, such events should not be ignored in resilience analysis.

Compound events serve as a major trigger for coupled disruptions. Driven by the concurrent hazards, it becomes highly possible to see a greater scale of disruptions or cascading failures across sectors.
The standard numerical tools for analyzing compound events is the correlation analysis and joint distributions. However, characterizing the occurrence likelihood and impacts of these events is a challenge.

\subsubsection{Drifting Mean Values}

This refers to a drift in the average level (normal/typical condition) towards the more extreme directions. Climate change may gradually shift the mean values and create new concerns. Reference~\cite{yalew2020impacts} conducted a review to summarize the climate impacts on renewable energy, hydropower, and bioenergy resources. 
Regional studies have also documented such impacts, including Norway~\cite{seljom2011modelling}, Eastern Africa~\cite{sridharan2019resilience}, and Brazil~\cite{de2010vulnerability}. 

In fact, drifting mean values have rich implications for energy infrastructure design and demand forecasting. As climate continues to evolve, it becomes essential to take this drifting effect into consideration.

\subsubsection{Connections and Differences}

There are empirical connections between the extremes and the mean values. Changes in long-term averages may increase the likelihood and intensity of extreme events. For example, rising average temperature are linked to frequent heatwaves, and sea level rise leads to coastal flooding risks. 
In addition, extreme prediction classically applies percentage changes on top of average values (baseline), where quantiles are assumed to be roughly proportional to averages. Similar strategies are widely applied to determine backup or reserve capacity for resilience.

However, extremes and mean values actually carry distinct implications for system resilience. Their differences lie in the resources to use, operating objectives, and mitigation strategies. In principle, the energy demand increment due to average changes should be handled by regular scheduling, while for extreme events, many resilience resources might be costly and rarely activated. Operating objectives for extreme events prioritize safety over average conditions.
As for mitigation strategies, failure and restorative actions are allowed for extreme events, which is never the case for normal conditions.

Resilience literature often emphasizes extreme events over mean-value shifts, but one must be cautious that an underestimated average drifts is also problematic and will take up the capacity reserve for resilience purpose.

Currently, our understanding of distribution shifts (especially extreme events) is far from adequate. These shifts may cause systematic biases and fail the classical trend extrapolation, so model validity check is necessary.
Unreliable estimation is another major difficulty. Reference~\cite{preston2016resilience} pointed out the varying surroundings and uncertainty that distribution shifts might bring to the energy systems. 

\subsection{Extreme Event Description} \label{subsec:cat-and-des}

Multi-energy systems are threatened by a variety of extreme events, which refer to abnormal weather conditions are at the most unusual extremes (e.g. below 10\% or beyond 90\% quantile~\cite{parmesan2022climate}) of the historical distribution. Interdisciplinary knowledge of climate science is essential to inform the description of extreme events.

Typical extreme events include hurricanes (or typhoons), ice storms (or snow storms, winter storms, blizzards), tornadoes, heat waves, wildfires, floods, tsunamis, and earthquakes. 

Different extreme events show huge disparities. For example, hurricanes are surface natural disasters, so underground transmission pipelines will not be affected. Earthquakes will affect all energy system components. Hurricanes often move fast~\cite{li2018minimax} but wildfires are located in one region. 

\begin{table*}[t] 
	\caption{Typical Extreme Events and The Major Characteristics}
	\label{tab:event-summary}
	\centering
	\setlength\tabcolsep{10pt}
	\begin{threeparttable} 
		\begin{tabular}{lp{8.5em}p{6em}lp{10.5em}p{14em}}
			\toprule
			Events     & Impact Region                & Duration                & Frequency           & Predictability             & Climate Change Impacts                                \\
			\midrule
			hurricane  & coastal regions              & 2+ weeks                & $\approx$2 per year    & moderate accuracy, days in advance   & risen intensity, nearly unchanged frequency                  \\
			tornado    & inland plains                & 20+ minutes             & $\approx$1,000 per year & bad accuracy, hours in advance   & more clustered, nearly unchanged frequency                   \\
			ice storms & high-latitude regions        & 3 hours to 5+ days      & 10--15 per year      & moderate accuracy, days in advance   & increased frequency                                   \\
			heat wave  & low-latitude regions         & 4 days on average       & $\approx$6 per year    & good accuracy, days to a week in advance   & risen intensity, longer duration, increased frequency \\
			wildfire   & inland regions (most)        & 1 hour to even weeks    & $\approx$1,300 per year & good accuracy, days in advance   & increased frequency                                   \\
			flood      & coastal and riparian regions  & 10 minutes to 2+ hours & 5--20 per year       & moderate accuracy, hours to days in advance   & risen intensity (most), more frequent and large flood   \\
			tsunami    & coastal regions              & 1 hour to days          & 0--2 per year        & moderate accuracy, minutes to hours in advance   & risen intensity                                       \\
			earthquake & tectonic plate edges (most)  & 10 seconds to minutes   & $\approx$16 per year   & bad accuracy, seconds to minutes in advance   & limited to no changes   \\
			\bottomrule                             
		\end{tabular}
		\begin{tablenotes}
			\item Note: This table only covers the most popular characteristics and their typical values, while special cases beyond the value ranges may surely exist. Data in the U.S. are used to show the typical duration and annual frequency. 
		\end{tablenotes}
	\end{threeparttable}
\end{table*} 

Describing the diverse characteristics of extreme events is generally a challenge, but classification could give an insight to show the similarities and differences. We extend the results in \cite{wang2015research} and further summarize the statistics in Table~\ref{tab:event-summary}.

In Table~\ref{tab:event-summary},
hurricanes are rare but destructive events in the U.S. that last for weeks in coastal regions, and tornadoes may happen more frequently in inland plains. We have good techniques to predict heat waves and wildfires, but tornadoes and earthquakes are far less predictable. Similar event categorization is developed in the existing effort~\cite{jasiunas2021energy}, with a unified model describing weather-driven threats in terms of speed, size, sustention, spread, and sureness.

Table~\ref{tab:event-summary} also summarizes how the event characteristics may evolve due to climate change, and almost all the events will experience a discernible change of intensity, frequency, and duration. These statements closely align with \cite{preston2016resilience}.

\subsection{Modeling and Prediction of Extreme Events} \label{subsec:mod-and-pred}

Extreme event modeling aims to use analytical models and historical data to investigate the abnormal weather conditions. 
There are two kinds of methodologies: model-based and data-driven approaches. The first kind relies on numerical simulations and physical laws, while the second kind extrapolates trends from historical data. Climate change is incorporated as coefficient variations or additive effects.
We next dive into the latest progress of extreme event simulation (model-based) and prediction (data-driven).

\subsubsection{Extreme Event Simulation}

Dynamic or time-series equations are formulated to simulate the future meteorological scenarios, and estimate both the chance of occurrence and the potential intensity. Two popular options are climate models and weather models. These models are very large-scale and complicated, but there are some existing full-fledged software and toolkit for practical use.

Climate models use physical equations to study the transfer, conversion, and interaction of energy and materials in climate systems~\cite{noaa2022climate}. These models often separate Earth surface into a three-dimensional grid of cells and the spatial resolution can be adjusted according to the practical needs. These models are powerful to answer what-if questions by simulating a series of scenarios of interest. In addition, climate models can account for the climate change effects endogenously because physical processes such as emission dynamics and heat transfer are already captured.

Climate models can be classified as either regional or global types. Three regional climate simulators were used in \cite{totschnig2017climate}, including the ARPEGE, the RegCM3, and the Remo model. As for global models, the coupled model intercomparison project (CMIP) might be the most well-known efforts starting in 1995. Currently, CMIP and its data infrastructure have already supported the IPCC and other international climate reports. CMIP3 and CMIP5 climate models were implemented in \cite{sridharan2019resilience} where the authors further adopted bias correction and spatial disaggregation to correct systematic errors.

Weather models or numerical weather prediction~(NWP) models employed a set of physical equations and numerical computation to model the atmosphere and its physical processes~\cite{noaa2022numerical}. Reference~\cite{chawla2018assessment} utilized a mesoscale NWP model WRF to analyze extreme rainfall events with sensitivity analysis. 
A climate model and a weather model were coordinated in \cite{hawcroft2021benefits} as an ensemble scheme.

It is natural to incorporate climate change impacts into extreme event simulators by endogenously changing the key parameters for scenario analysis. These fluctuations may be non-intuitive and rely heavily on expert knowledge. Capturing uncertainty in future impacts and risk propagation is also critical in such simulations.

\subsubsection{Extreme Event Prediction}

Climate and weather models are computationally expensive due to their complex physics. Differently, extreme event prediction is driven by data rather than physic laws. It applies unstructured models to directly learn data correlations for trend extrapolation.

Data especially high-quality data sources are key elements in extreme event prediction. A popular resource is the reanalysis data, which are generated using an atmospheric reanalysis for data assimilate and quality control. Past observations will be combined to create physically consistent time-series data.

MERRA-2 is a well-known reanalysis dataset managed by the National Aeronautics and Space Administration (NASA). Reference~\cite{jamieson2020quantification} tuned the MERRA-2 data to involve geographical downscaling and wind speed correction according to tower and turbine heights. Reference~\cite{fu2017integrated} ran the extreme wind simulator to estimate the wind speed and intensity using the ERA-Interim reanalysis data. 
IWEC is another meteorological database dedicated to building energy simulation, whose original source came from the National Climatic Data Center. Reference~\cite{ascione2017resilience} derived extreme scenarios from IWEC and interpolated it to build a eight-year-long hourly dataset.

Extreme event prediction typically uses statistical or machine learning models. For the first kind, the generalized extreme value (GEV) distribution is a popular option:
\begin{subequations}
	\begin{align}
		& \mathsf{CDF}_\mathit{GEV} (x) = e^{-h(x)} \\
		& h(x) = 
		\begin{cases}
			(1 + \xi (x - \mu) / \sigma)^{-1/\xi}, & \xi \neq 0 \\
			e^{-(x - \mu) / \sigma}, & \xi = 0
		\end{cases}
	\end{align}
\end{subequations}
where $x$ is a weather variable, $\mu$ is the mean value, $\sigma$ is the standard deviation, and $\xi$ is the shape parameter. 

Reference~\cite{liu2020data} analyzed the historical hurricane statistics to establish uncertainty intervals for hurricane tracks. Reference~\cite{finkel2023revealing} conducted long-term climatological risk assessment by combining different short-term trajectories to tighten the uncertainty bounds of long-term estimations.

Recently, learning-based models have gained strong interest for extreme event prediction.
In \cite{lam2023learning}, the reanalysis data were fed into a machine learning model to improve the tropical cyclone tracking and extreme temperatures. In \cite{chattopadhyay2020analog}, a capsule neural network was proposed to predict the likelihood of extreme cold and heat waves at an accuracy up to 88\%. Reference~\cite{kim2019deep} predicted hurricane tracks with a convolutional LSTM network capturing spatiotemporal properties. 

Beyond the above discussions, some studies have accounted for distribution shifts.
Unlike extreme event simulation, the prediction approaches is only able to consider distribution shifts in an exogenous fashion. They often assume an additive form and the task is to estimate the baseline states (drifting mean values) and the additive residuals separately. 

\subsection{Impacts of Extreme Events on System Components} \label{subsec:imp-sys-comp}

This subsection studies how to characterize the extreme event impacts on the components of multi-energy systems. Here, the vulnerable components may belong to any sectors (electricity, natural gas, heat, or others) and any sections (generation, transmission, distribution, or demand). 

\begin{figure}[t]
	\centering
	\includegraphics[width=0.42\textwidth]{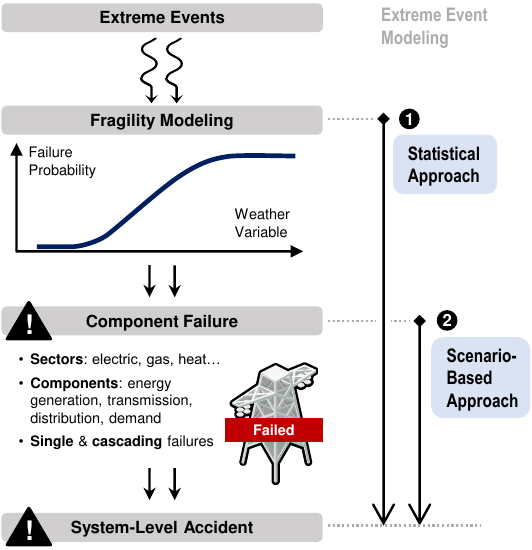}
	\caption{Two approaches to analyze the extreme event impacts on system components. Difference lies in the settings of component failure.}
	\label{fig:extreme-event-impact}
\end{figure}

The main focus is impact modeling and analysis, covering statistical and scenario-based approaches and their respective advantages and limitations.
Fig.~\ref{fig:extreme-event-impact} showcases a classical flowchart and two approaches for impact analysis. Statistical approaches clearly have better tracking of climate change impacts and hold greater potential for integrating interdisciplinary knowledge from climate science.

\subsubsection{Statistical Impact Analysis}

The idea is to establish a probabilistic mapping between weather observations and component failures. A common assume is that climate change may affect the frequency and intensity of extreme events, but has little impacts on these mappings.

Classical statistical approaches include fragility functions and failure rate functions.

A fragility function relates component failure probability to weather conditions.
It can be visualized with a fragility curve, as shown in Fig.~\ref{fig:key-concepts}(b). Calibrating a fragility function is a regression task: Fit the meteorological data and failure rate statistics by least squares or maximum likelihood estimators. 

Regressor selection depends on the data and application. In the literature, the top three options are normal, log-normal, and piecewise linear functions, expressed as follows:  
\begin{subequations}
	\begin{align}
		& \lambda (x) = \mathsf{CDF}_\mathcal{N} (x) = \Phi \! \left( \frac{x - \mu}{\sigma} \right) \\
		& \lambda (x) = \mathsf{CDF}_\mathit{LogNorm} (x) = \Phi \! \left( \frac{\ln(x) - \mu}{\sigma} \right) \\
		& \lambda (x) = \mathsf{CDF}_\mathit{PWL} (x) = \! 
		\begin{cases}
			0, & x < x_\text{min} \\
			\frac{x - x_\text{min}}{x_\text{max} - x_\text{min}}, & x_\text{min} \le x \le x_\text{max} \\
			1, & x > x_\text{max}
		\end{cases}
	\end{align}
\end{subequations}
where $x$ is a weather variable, $\mu$ and $\sigma$ are the mean and standard deviation, and $\Phi (\cdot)$ is the cumulative distribution function of the standard normal distribution.

Reference~\cite{yang2022resilience} collected several hurricane-fragility models for distribution lines, towers, substations, and wind turbines. A log-normal fragility curve was developed in \cite{zhang2020resilience} to describe the failure rates of transmission towers and conductors under hurricane attacks. There are many other useful distributions, such as Pearson for modeling extreme wind droughts~\cite{potisomporn2024extreme}. 


A failure rate function, however, puts more emphasis on the performance degradation from a baseline rate (normal status). For example, the failure rate of hurricane-attacked transmission lines was modeled in the following way~\cite{wang2016resilience}:
\begin{align}
	\lambda (x) = \left( 1 + \beta h(x) \right) \lambda
\end{align}
where $\lambda$ is the baseline failure rate, $\beta$ is a weighting factor, and $h(x)$ is a given function (e.g. quadratic, polynomials). 


Another kind of complicated statistical models could be formulated by the connections between components. One of the classical expressions is shown as follows:
\begin{align}
	\lambda (x) = 1 - \prod\nolimits_i (1 - \lambda_i)
\end{align}
where the fragile component is assumed to be connected to several other components that are indexed by $i$, and these components might break down independently or jointly. 

Such relationships are widespread in networked energy systems. 
For example, transmission line faults are directly linked to tower faults, as each line connects two towers. Similarly, compressor station failures can affect gas pipelines.

Advanced fragility functions are commonly composite and multivariate.
A typical case is that chilly wind and freezing temperature may collectively increase the chance of a frozen gas pipeline. 
In \cite{zhang2020resilience}, a complex model combined the influences of strong wind and heavy rainfall into an equivalent wind speed and then applied the existing empirical formula. Reference~\cite{bennett2021extending} extended fragility curves to consider the remaining operable capacity after hurricanes as well.

Further, the fragile models or curves may change over time due to climate-induced distribution shifts. Common solutions include constructing conservative distributions/functions or incorporating higher-quantile thresholds in the analysis.

\subsubsection{Scenario-Based Impact Analysis}

The analysis prescribes extreme events under certain fault assumptions. This paradigm offers a unified treatment of extreme events but depends on effective assumptions and reasonable balance between resilience and redundancy.

A typical example is the N-k contingency which sets an extreme scenario that any k components fail at the same time. N-k contingencies are conservative because it completely ignores the potential dependence between different failed components. Reference~\cite{li2020multi} further constructed a temporal-spatial natural disaster destruction model, involving a group of matrices to capture the hurricane trajectories. Reference~\cite{zhang2022resilience} applied the complex network theory to identify the extreme scenarios of peak energy consumption and line faults according to the betweenness and connection properties. Reference~\cite{fu2017typical} generated scenarios of air temperature and photovoltaic outputs using Latin hypercube sampling and copula functions. 

Most scenario-based approaches, however, rarely account for the coupled disruptions that may occur in multiple sectors. Existing works are mainly focused on varying levels of disruption severity in a single sector.

\section{Multi-Energy Coordination and Resilience Resources} \label{sec:resource}

This section reviews the multi-energy coordination patterns and the potential pool of resilience resources. 

\subsection{Evolution of Multi-Energy System Models} \label{subsec:evol-mes-model}


\subsubsection{Energy Conversion}

\begin{figure}[t]
	\centering
	\includegraphics[width=0.45\textwidth]{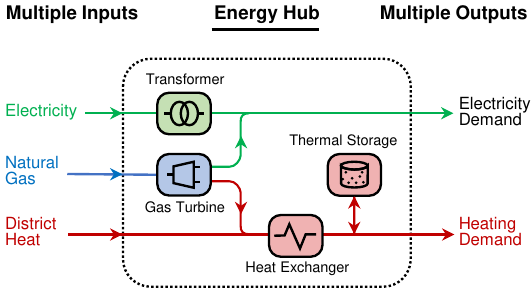}
	\caption{An energy hub with three inputs and two outputs.}
	\label{fig:energy-hub}
\end{figure}

Energy hub is a standard and widely-used model for multi-energy systems~\cite{mohammadi2017energy}. As shown in Fig.~\ref{fig:energy-hub}, an energy hub describes the energy conversion via a two-port network that typically involves multiple inputs and multiple outputs. 

An energy hub model assigns conversion rates to different input-output pairs, shown as follows:
\begin{align}
	P_\text{out} = H \, P_\text{in} - S \, P_\text{es} 
	\label{eqn:vinvout}
\end{align}
where $P_\text{out}$/$P_\text{in}$ are the energy output/input vectors; $H$ and $S$ are two coupling matrices for conversation rates; $P_\text{es}$ is the power of energy storage.

There are multiple extensions in the literature. A direction is bidirectional conversion, and a clean solution is to use two hubs in parallel with separate conversion rates and internal conversion paths. Another direction is to consider nonlinear/dynamic conversion rates, and the nonlinearity can be captured by machine learning or sequential approximation. For instance, an electrolyzer model with dynamic conversion efficiency is given below~\cite{gebreslasie2023modeling}. 
\begin{subequations}
	\begin{align}
		& P_{\text{H}_2} = \eta_\text{st} P_\text{elec} \\
		& \eta_\text{st} = 0.475 M_{\text{H}_2} \mathit{HHV} / V_\text{st}
	\end{align}
	\label{eqn:h2-electrolyzer}
\end{subequations}
where $P_{\text{H}_2}$ and $P_\text{elec}$ are the power of hydrogen and electricity; $\eta_\text{st}$ is the dynamic conversion efficiency that is proportional to the hydrogen molar mass $M_{\text{H}_2}$ and the hydrogen heat value $\mathit{HHV}$ but inversely proportional to the stack voltage $V_\text{st}$. The conversion rate differs when the operating voltage changes.

Feasible region model is the second classical option in multi-energy systems. It has been widely applied for combined heat and power~(CHP) generators (convert fuels into electricity and heat) and many other variations~\cite{chen2014increasing}. Following Fig.~\ref{fig:chp-region}, the operation of a CHP generator is often modeled by:
\begin{subequations}
	\begin{align}
		& V_\text{out} = [ P_\text{elec}, P_\text{heat} ]^\top \in \Omega_\text{CHP} \\
		& \Omega_\text{CHP}: \ \alpha_m P_\text{elec} + \beta_m P_\text{heat} \le \gamma_m, \quad \forall m
	\end{align}
\end{subequations}
Here, the electricity and heat power outputs must satisfy a set of linear constraints.

\begin{figure}[t]
	\centering
	\includegraphics[width=0.45\textwidth]{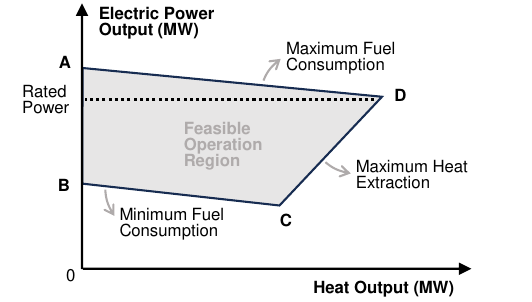}
	\caption{A feasible operation region for a combined heat and power generator.}
	\label{fig:chp-region}
\end{figure}

\subsubsection{Energy Delivery}

Energy bus is a simple and generic energy-flow-based model for multi-energy delivery. 
This model resembles energy hubs, with a two-port network capturing all the flows between different energy sectors. It might perform poorly for ignoring specific operating constraints.

Apart from energy bus models, there are specialized operating models arising from different sectors. 

For electric power grids, linearized power flow models such as the direct current (DC) power flow model have been widely applied in the literature and in practice. It well captures active power but ignores reactive power, nodal voltage amplitude, and network loss.
Its formulation is expressed below:
\begin{subequations}
	\label{eqn:dcpf}
	\begin{align}
		& P_j^\text{inj} = \sum_{k: j \rightarrow k} P_{jk} - \sum_{i: i \rightarrow j} P_{ij}, \quad \forall j \label{eqn:dcpf-1} \\
		& P_{ij} = -b_{ij} ( \theta_i - \theta_j ), \quad \forall (i, j)
	\end{align}
\end{subequations}
where $P_j^\text{inj}$ is the injection active power at the node $j$; the right hand side of (\ref{eqn:dcpf-1}) involves all the branch power flow leaving and injecting into the node; $P_{ij}$ is the branch power flow between the node $i$ and $j$; $\theta_i$ is the voltage phase angle; and $b_{ij}$ stands for the branch susceptance.


Distflow model and several variations have emerged~\cite{farivar2013branch} to address the drawback of DC power flow by enabling the computation of reactive power flows, nodal voltage amplitude, and network losses. This model is known to be exact in radial grids and computationally efficient (second-order cone program), so it serves as the standard tool for distribution grids. The detailed formulation is formulated as:
\begin{subequations}
	\begin{align}
		& P_j^\text{inj} = \sum_{k: j \rightarrow k} P_{jk} - \sum_{i: i \rightarrow j} (P_{ij} - r_{ij} L_{ij}), \quad \forall j \\
		& Q_j^\text{inj} = \sum_{k: j \rightarrow k} Q_{jk} - \sum_{i: i \rightarrow j} (Q_{ij} - x_{ij} L_{ij}), \quad \forall j \\
		& U_i - U_j = 2 ( r_{ij} P_{ij} + x_{ij} Q_{ij} ) \notag \\
		& \qquad \qquad \ - ( r_{ij}^2 + x_{ij}^2 ) L_{ij}, \quad \forall (i, j) \\
		& L_{ij} U_i \ge P_{ij}^2 + Q_{ij}^2, \quad \forall (i, j ) \label{eqn-distflow-4}
	\end{align}
\end{subequations}
where $P$ and $Q$ are symbols denoting the active and reactive power; $L$ and $U$ are denoting the squared current and voltage magnitude; $r_{ij}$ and $x_{ij}$ are the resistance and reactance values of the branch $i$--$j$.


For natural gas or hydrogen pipeline networks, the classical steady-state gas flow model is nonlinear with respect to the pipeline volumetric flow and nodal pressure:
\begin{subequations}
	\label{eqn:gas-flow}
	\begin{align}
		& G_j^\text{inj} = \sum_{k: j \rightarrow k} G_{jk} - \sum_{i: i \rightarrow j} G_{ij}, \quad \forall j \\
		& \Pi_i^2 - \Pi_j^2 = \sigma_{ij} G_{ij} | G_{ij} |, \quad \forall (i, j)
	\end{align}
\end{subequations}
where $G_j^\text{inj}$ is the injection at the node $j$; $G_{jk}$ and $G_{ij}$ are the branch flows leaving and injecting into the node $j$; $\Pi_i$ is the nodal pressure; and $\sigma_{ij}$ is a resistance factor.


For district heating networks, the classical steady-state water flow model is nonlinear with respect to the branch volumetric flow in pipelines:
\begin{subequations}
	\begin{align}
		& M_j^\text{inj} T_j^\text{inj} = \sum_{k: j \rightarrow k} M_{jk} T_j - \sum_{i: i \rightarrow j} M_{ij} T_j, \quad \forall j \\
		& T_j = T_\text{amb} + (T_i - T_\text{amb}) \exp(-\kappa_{ij} / M_{ij}), \ \forall (i, j)
	\end{align}
\end{subequations}
where $M_j^\text{inj}$ is the injected water flow with a temperature of $T_j^\text{inj}$ at the node $j$; $M_{jk}$ and $M_{ij}$ are the branch water flows leaving and injecting into the node $j$; $T_j$ is the temperature at the node $j$; $T_\text{amb}$ is the ambient temperature that causes the heat loss along the delivery; and $\kappa_{ij}$ is a conversion factor.


\subsubsection{Energy Storage}

Linearized energy storage model is commonly applied in the literature. Resources with similar temporal dynamics could serve as the so-called generalized or virtual energy storage~\cite{barala2021virtual}, such as time-shiftable loads, natural gas pipelines (storage effect), and smart buildings (thermal inertia). The model is expressed as follows:
\begin{subequations}
	\begin{align}
		& P_t^\text{es} = P_t^\text{chg} - P_t^\text{dchg}, \quad \forall t \\
		& 0 \le P_t^\text{chg} \le \overline{P}_\text{es} u_t, \quad \forall t \\
		& 0 \le P_t^\text{dchg} \le \overline{P}_\text{es} (1 - u_t), \quad \forall t \\
		& \mathit{SoC}_{t+1} = \mathit{SoC}_t + E_\text{es}^{-1} \left( \eta P_t^\text{chg} - P_t^\text{dchg} / \eta \right) \Delta t, \forall t \label{eqn:storage-4}
	\end{align}
\label{eqn:storage}
\end{subequations}
where $P_t^\text{es}$ is the energy storage output at the time step $t$ with positive values denoting charging and negative values denoting discharging; $P_t^\text{chg}$ and $P_t^\text{dchg}$ are two auxiliary variables for charging or discharging; $u_t$ is a binary variable to decide either charging or discharging; $\mathit{SoC}_t$ is the state of charge; $\eta$ is an efficiency coefficient, $E_\text{es}$ is the total battery capacity, and $\Delta t$ is the duration of a time step.


Linear models can hardly capture nonlinear characteristics such as degradation and dynamic charging efficiency, but recent advances incorporate these complexities into more sophisticated energy storage models.

\subsubsection{Support Systems}

For transportation and water systems, the commodity flow model is a popular option:
\begin{align}
	f_{j,m}^{\text{inj}} = \sum_{k: j \to k} f_{jk,m} - \sum_{i: i \to j} f_{ij,m}
\end{align}
where $f_{j,m}^{\text{inj}}$ is the injected flow of traffic/waterway/channel $m$; $f_{jk,m}$ and $f_{ij,m}$ are the branch flows transporting inward and outward through the network without losses.  

There are advanced models in the literature. Transportation systems can be modeled by spatio-temporal network diagrams~\cite{he2024resilient} or graph-based systems. Water management can consider hydraulic attributes and pipeline dynamics~\cite{yu2023resilience}.

\subsection{Potential Resilience Resources} \label{subsec:res-resource}

Multi-energy systems have a large pool of resilience resources from different sectors. 
Table~\ref{tab:res-resources} collects a list of resilience resources coming from the energy sources, networks, load, and storage side. 

\begin{table}[t] 
	\caption{Typical resilience resources in a multi-energy system}
	\label{tab:res-resources}
	\centering
	\begin{threeparttable} 
		\begin{tabular}{l p{21em}}
			\toprule
			Category & Typical Examples \\
			\midrule
			energy supply & traditional generators (spinning and non-spinning reserve), CHP generators, solar panels, wind turbances, diesel generators, black-start generators \\
			energy networks & redundant energy transmission lines, controllable network topology, controllable network impedance, static VAR compensators, natural gas compressors \\
			energy demand & uninterruptible power supply, backup generators, multi-energy buildings, electric vehicles, demand response, hydrogen electrolysis \\
			energy storage & pumped hydro storage, battery storage, fuel cells, natural gas / hydrogen storage tanks, thermal energy storage, mobile energy storage systems \\
			\bottomrule
		\end{tabular}
	\end{threeparttable}
\end{table} 

\subsubsection{Resources from Energy Supply}

Traditional generators such as fossil fuel units are the backbone for securing a desired level of resilience. Weather-driven generators may degrade during natural disasters, and this truly happened in the Texas winter storm 2021~\cite{utaus2021timeline} when wind farms ran out of services because of insufficient winterization.

CHP generators belong to flexibility-limiting resources. Inside, the electricity and heating outputs of CHP generators are coupled and cannot be separately regulated. It is possible to expand the flexibility of a CHP generator using electric or thermal energy storage facilities.

Distributed generators such as utility solar panels and wind turbines are local resources at the grid edge. During a disaster, these generators are able to reliably support local recovery. Reference~\cite{gupta2019achieving} evaluated how the solar photovoltaic systems and smart batteries could support a local community of 82 households. According to \cite{poudel2022operational}, small modular reactors were also able to provide flexible electric and heating supply.

In addition, typical resources consists of all generators with black-start capability (e.g. diesel generators, batteries).

\subsubsection{Resources from Energy Networks}

Energy transmission networks are among the hardest-hit components in the entire system, so the redundant capacity in these networks become an effective mitigation action. Electric power grids often meet the N-1 or even N-k conditions by promoting loop topology and backup transmission corridors. Multi-energy systems further improve the resilience performance by activating a larger pool of resilience resources from different energy sectors.

Flexible topology is a resilience resource on the network side. A well-known example is microgrid formation, which lets microgrids under disruptions shift to island mode and keep local energy balancing. Reference~\cite{hussain2019microgrids} reviewed the use of microgrid as a resilience resource through microgrid formation, networked microgrids, and dynamic microgrids. 

Flexible network impedance is a potential resilience resource as well. This controllable resource is not very common and might be small-scaled in real-world systems. Reference~\cite{song2022convex} found that a flexible-impedance transmission line could be modeled as a constant-impedance line linking with a pair of adjustable transformers. 

System inertia and inertia-based coordination serve as a special kind of resilience resources. Here, the slow inertia in natural gas or district heating pipelines could be utilized to mitigate the fast fluctuations and address the short-term imbalance in electric power grids. 

In addition to the above resources, there are some additional network-side facilities, including natural gas compressors and reactive power compensators.

\subsubsection{Resources from Energy Demand}

Two critical resources from the demand side are the uninterruptible power supply (UPS)~\cite{ferraro2020uninterruptible} and backup generators (often fueled by natural gas, propane, or diesel). They are often small-scaled energy supplier owned by the customers. 

Multi-energy buildings act as a potential resilience resource because their energy use can flexibly track coordination signals. Such a building was proposed in \cite{mehrjerdi2020zero} with boosted energy resilience. The hydro, wind, solar, and hydrogen energy were coordinated to handle different failure patterns.

Electric vehicles (EVs) contribute to household energy security as mobile resilience resources. Smart EV charging make their energy use adaptive, and the mobility offers additional spatial flexibility. Reference~\cite{hussain2022resilience} summarized the practice of using EV batteries (first- and second-life) as a resilience resource during different phases of energy outages. Reference~\cite{liu2025leveraging} proposed a vehicle-to-home scheme that enabled hybrid EVs to power residential units during grid outage.

Broadly, a diverse category of demand response (DR) resources are eligible to provide resilience services. In \cite{clegg2019integrated}, a demand response program from natural gas systems had a special role under extreme disturbances.

Hydrogen electrolyser is a novel and fully controllable electric demand, and it is flexible as a resilience resource. Reference~\cite{dozein2021fast} proposed a dynamic model to capture the fast frequency response provided by hydrogen electrolysers, and then validated the value of grid-scale electrolysers in supporting the resilience of future energy systems.

\subsubsection{Resources from Energy Storage}

Various types of energy storage facilities are crucial flexibility provider for multi-energy systems, and these facilities include pumped hydro energy storage, battery energy storage, natural gas storage tanks, and thermal energy storage. Note that many energy storage models require interdisciplinary knowledge.

Optimal installation locations and capacity sizing of batteries becomes a potential resource for resilience enhancement. The urban-scale resilience issue was studied in \cite{ping2025city} and the cross-domain data of geographic information systems were integrated in the decision as well.

Another resource is the shared energy storage~\cite{khojasteh2025distributed}, which could expand the storage  within an energy community by sharing idle capacity and storing or releasing excess energy.

Mobile energy storage stands for any storage facility with mobility characteristics. A typical case is to use a truck to carry battery storage units for local energy supply in emergency. For instance, reference~\cite{mehrjerdi2021resilience} demonstrated the role of mobile battery storage for enhancing resilience, improving self-adequacy, and cutting down the load shedding and operation cost. 

Long-term energy storage enhances energy resilience by bridging prolonged supply-demand gaps. Typical technologies include hydrogen storage, pumped hydro storage, compressed air storage, and thermal storage. In \cite{harsini2023resilience}, multi-level energy storage system including seasonal (storage hydropower) and hourly storage (super-capacitor) was integrated. 


\subsubsection{Further Discussions}


There is an obvious overlap between resilience and flexibility resources, but they have distinct differences in the practical applications. In general, flexibility resources are just part of resilience resources, and there are many other expensive backups that might never operate in years/months, such as black-start generators and uninterruptible power supply. 

Resilience resources can be influenced by climate change as well. In particular, the intensified extreme events may hit the connecting facilities and cause facility failures, which will cut off the interactions between sectors. The drifting mean values may change the amount of resources and limit the ability of mutual support. For instance, a reduction in rainfall leads to less water availability, and thus limits the operation of electrolyzers for grid support.

\subsection{Resilience Enhancement Strategies} \label{subsec:res-enh-strategy}

\begin{figure}[t]
	\centering
	\includegraphics[width=0.42\textwidth]{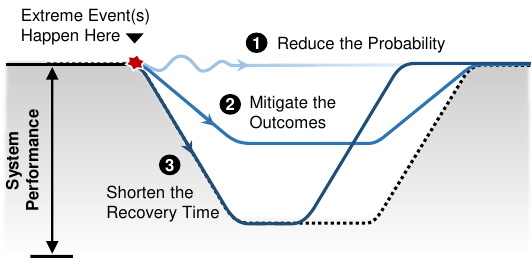}
	\caption{Three typical resilience enhancement strategies. The dotted line is the benchmark case and three different enhancement strategies are compared.}
	\label{fig:3-strategies}
\end{figure}

There are three effective solutions to improve system resilience: reduce the probability of failure, mitigate the possible outcomes, and shorten the recovery time. 
Fig.~\ref{fig:3-strategies} graphically illustrates these solutions by three colored trajectories that are compared with the original resilience curve.

Investing in new facilities is perhaps the most direct and effective way to improve system resilience. This strategy expands the list of resilience resources against external disruptions and reduces the likelihood of system failures. Equipment management and enhancement, such as transmission line upgrade and dynamic line rating, is able to greatly reduce the failure probability and accelerate the recovery procedures. It is often a critical step to identify the bottlenecks of resilience performance before taking any customized actions. 

Energy flow optimization is a special strategy to maximize the synergy between energy sectors. Building upon the cross-sector coordination, it is beneficial to optimize the connection and conversion between different energy hubs for topological resilience. Reference~\cite{ma2018optimal} made the structural planning for multi-energy systems by optimizing the selection and connections of system devices. Reference~\cite{salimi2020replacement} discussed a replacement strategy to increase the resilience against seismic waves. Specifically, this strategy shifted energy delivery pathways with less involvement of natural gas pipelines.

Inertia-based coordination explores the temporal flexibility and leverages the slow pipeline inertia to smooth the interruptions in electric power grids. For instance, in \cite{huang2021resilience}, the thermal inertia of heat/cooling loads were used to improve resilience for a urban multi-energy system. 

Pre-heating/cooling and charging control shares a similar idea to adopt temporal flexibility. A typical case is the pre-cooling of buildings during heat waves~\cite{zeng2022investigation}. A strategy was developed to pre-cool the building during off-peak hours to minimize the risks of overhating and cooling capacity shortfall. In \cite{hussain2022resilience}, several EV-involved strategies for resilience enhancement were discussed, including the off-grid charging and selective energy allocation to EVs.

Network splitting and microgrid formation are widely applied in local restoration after outages, which largely reduce outage consequences and recovery time. Reference~\cite{hussain2019microgrids} compared and discussed four resilience-oriented strategies including the proactive scheduling, outage management, feasible islanding, and operation against disruptions. Reference~\cite{ma2025optimal} proposed a unified scheduling model with both normal and islanding constraints to reduce the demand loss in worst cases. Reference~\cite{ghasemi2025improving} networked and coordinated multiple multi-energy microgrids to secure the system during sudden outages in upstream gas and electricity networks.

Dynamic network topology is another way to use spatial flexibility, but the scheduling of topological shifts might be complicated. This issue was discussed in \cite{xu2019resilience} for an integrated electricity-transportation system.

Pre-positioning mobile resources deploys spatial flexibility to find the most effective allocation and routine of resilience resources for critical regions. This strategy can reduce the adverse outcomes and accelerate the recovery process. Reference~\cite{gharehveran2023resilience} proposed a bi-level model to pre-position vehicle-mounted energy storage for emergence. Reference~\cite{wang2021scheduling} made a further improvement to allow separation of the mobile energy storage and the carrier vehicles. 

Any single strategy is rarely sufficient to achieve desired resilience. Instead, resilience can be improved through resource diversification and prioritization, with multiple strategies deployed in an appropriate order.

A diverse and priority-driven allocation of resilience resources has great potential to enhance the temporal and spatial flexibility simultaneously. Reference~\cite{chen2017modernizing} discussed many practical issues including cold load pickup, distributed data collection, and peer-to-peer communication.

\section{Resilience Evaluation of Multi-Energy Systems} \label{sec:evaluation}

This section proposes a hierarchical structure to categorize evaluation metrics and multi-metric approaches.

\subsection{Categories and Metric Design} \label{subsec:cat-metric-design}

The existing evaluation approaches can be broadly categorized according to the following aspects~\cite{shandiz2020resilience}:
\begin{itemize}
	\item Sectors/Subsystems: electricity, natural gas, heating, hydrogen, transportation, water, or any combinations.
	\item Resilience Phase: normal phase, disturbance phase, degradation phase, recovery phase, any combinations or the entire disturbance duration.
	\item Resilience Approach: qualitative or quantitative, deterministic or probabilistic.
\end{itemize}

%

Quantitative approach is still the major choice among all.
We next outline the key characteristics of quantitative metrics and summarize the main idea of metric design. This high-level discussion will be useful and informative to understand different evaluation metrics.

Evaluation metrics can be broadly demonstrated through the measures of levels, differences, or cumulative differences. 
They can also be categorized depending on whether they encompass the (internal) structural characteristics or (external) operational impacts.

%

\subsection{Component-Level Vulnerability Analysis Metrics} \label{subsec:com-level-metric}

The component-level vulnerability analysis is valid for a network node, a specific device (e.g., substation, gas pipeline), a sub-region, or an energy subsystem. 

Here are several widely used concepts for resilience evaluation. Let $R$ denote a resilience metric and $\mathit{MoP}$ denote a measure of performance, e.g. served demand, percentage of served demand, and thermal comfort. We apply these two symbols in different use cases: $R^\text{comp}_i$ refers to the resilience of component $i$, $R^\text{sys}$ refers to the system resilience, $\mathit{MoP}_i$ refers to the residual performance without component $i$, and $\mathit{MoP} (t)$ refers to the performance at time period $t$.
We also add failure rate $\lambda$, value of loss load $\mathit{VoLL}$ and expected energy not served $\mathit{EENS}$ (or demand not supplied).


\subsubsection{Failure-Related Metrics}

The first kind of metrics are simply expressed by failure rates, shown as follows:
\begin{subequations}
	\begin{align}
		& R^\text{comp}_i = \lambda_i \label{eqn:Relem1a} \\
		& R^\text{comp}_i = - \lambda_i \ln( \lambda_i ) \label{eqn:Relem1b}
	\end{align}
\end{subequations}
where $R^\text{comp}_i$ is the resilience measurement of component $i$, and $\lambda_i$ is the failure rate. Both equations (\ref{eqn:Relem1a}) and (\ref{eqn:Relem1b}) are an inherent property of components, and (\ref{eqn:Relem1b}) is typically expressed by information entropy.

The first kind is simple but useful. Reference~\cite{he2018robust} assigned a failure probability to each power line and gas pipeline, and then used the negative logarithm of failure probability as the resilience metric. 

\subsubsection{Structural Metrics}

The second kind is related to the structural or statistical property. A typical case is the following nodal degree metric:
\begin{align}
	\label{eqn:Relem4}
	R^\text{comp}_i = \deg(i)
\end{align}
where $\deg(\cdot)$ is the nodal degree, which is defined as the number of edges connected to a node. It is a useful statistical metric to measure the nodal connectivity that should include all coupled sectors and all energy dependence.

Reference~\cite{kumar2022resilience} analyzed the network topology by randomly removing some nodes and edges. In this setting, a greater value of (\ref{eqn:Relem4}) implied a potentially larger impacts.

\subsubsection{System-Oriented Metrics}

The third kind is formulated from the measures of system resilience. The status of a component is typically quantified by its contribution to the system resilience. 
Below are three typical expressions:
\begin{subequations}
	\begin{align}
		& R^\text{comp}_i = \mathit{MoP}_i \label{eqn:Relem2a} \\
		& R^\text{comp}_i = \mathit{MoP}_\text{\!max} - \mathit{MoP}_i \label{eqn:Relem2b} \\
		& R^\text{comp}_i = \frac{\mathit{MoP}_\text{\!max} - \mathit{MoP}_i}{\mathit{MoP}_\text{\!max}} \times 100\% \label{eqn:Relem2c}
	\end{align}
\end{subequations}
where $\mathit{MoP}_i$ is the remaining performance when component $i$ is offline, and $\mathit{MoP}_\text{\!max}$ is the maximal system performance but often set as the original performance in normal phase. It is evident that (\ref{eqn:Relem2a}) is a level-based metric, while the other two (\ref{eqn:Relem2b})(\ref{eqn:Relem2c}) are difference-based metrics.

The resilience curves (measure-of-performance curves) can be constructed in two ways: separate curves (mainstream) and unified curves (rare). It is common to formulate the resilience curves for different sectors separately and apply synchronization later in the composite evaluation process. 

This kind is very popular in vulnerability analysis. For example, the critical component identification was conducted in~\cite{jiang2022resilience} using a component-level index like (\ref{eqn:Relem2b}). The resilience enhancement strategy was to prioritize repairing the most vulnerable components one after another. Reference~\cite{bao2020modeling} used the nodal curtailment of electricity and natural gas demand as the evaluation metric, showing a similar expression of (\ref{eqn:Relem2c}). It was found in \cite{bao2020modeling} that this metric was informative to detect the spatial performance changes under wind storms. Component criticality ranking was then adopted to identify the vulnerable substations and generating units.

\subsubsection{Cost-Oriented Metrics}

The fourth kind is cost-oriented and they can be derived by extending (\ref{eqn:Relem2b}):
\begin{align}
	\label{eqn:Relem3}
	R^\text{comp}_i = ( \mathit{MoP}_\text{\!max} - \mathit{MoP}_i ) \; \mathit{VoLL}
\end{align}
where $\mathit{VoLL}$ is the value of loss load, as previously defined. The $\mathit{MoP}$ should use the quantity of energy load so that the first term in parentheses has a similar meaning of $\mathit{EENS}$. It is obvious that (\ref{eqn:Relem3}) is simply transforming the unserved load into an economic loss.

Reference~\cite{moslehi2018sustainability} developed a metric based on the normalized imposed cost~(\ref{eqn:Relem3}). The metric was declared to efficiently identify critical components of an integrated energy system.

\subsubsection{Probabilistic Metrics}

The final kind is the probabilistic versions of all the above metrics. This can be done by lifting with a new dimension of scenarios and computing a stochastic measure such as expectation or conditional value-at-risk. The formulations are given below:
\begin{subequations}
	\begin{align}
		& R^\text{comp}_i = \mathsf{E}_w [R^\text{comp}_{iw}] \label{eqn:Relem5a} \\
		& R^\text{comp}_i = \mathsf{CVaR}_\alpha [R^\text{comp}_{iw}] \label{eqn:Relem5b}
	\end{align}
\end{subequations}
where $\mathsf{E}_w [\cdot]$ is an expectation function over all scenarios $w$, and $\mathsf{CVaR}_\alpha [\cdot]$ is a conditional value-at-risk function parameterized by the probability threshold $\alpha$.

\subsection{System-Level Resilience Evaluation Metrics} \label{subsec:sys-level-metric}

The system-level evaluation studies the whole system ability to withstand external risks. This line of research has attracted a wide range of efforts~\cite{ahmadi2021frameworks} but extra efforts are required to summarize the existing metrics in a unified way.

\begin{figure}[t]
	\centering
	\includegraphics[width=0.42\textwidth]{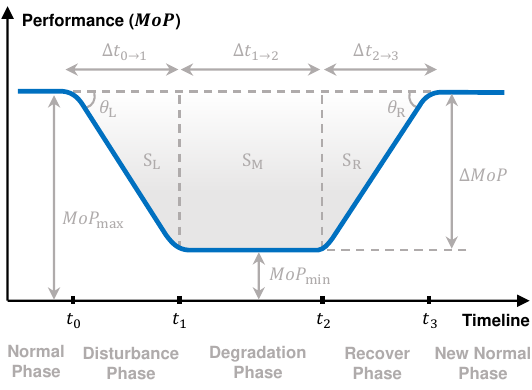}
	\caption{A resilience curve with timeline and measure of performance as the axes. Typical marks are assigned to the distances, areas, and angles that are useful for metric designs.}
	\label{fig:res-curve}
\end{figure}

A clear observation is that most system-level metrics are strongly relevant to the resilience curves. As shown in Fig.~\ref{fig:res-curve}, the y-axis is measured by the previous $\mathit{MoP}$, which are often set as the total energy demand or a single kind of demand (consider multiple curves). Several distances, areas, and angles of interest are marked accordingly in this figure. 

\subsubsection{Distance-Based Metrics}

The first kind of system metrics focus on the measure of differences and apply the following distance-based expressions:
\begin{subequations}
	\label{eqn-dist-mop}
	\begin{align}
		& R^\text{sys} = \mathit{MoP} (t) \label{eqn:Rsys1a} \\
		& R^\text{sys} = \mathit{MoP}_\text{\!max} - \mathit{MoP}_\text{\!min} \label{eqn:Rsys1b} \\
		& R^\text{sys} = \frac{\mathit{MoP}_\text{\!max} - \mathit{MoP}_\text{\!min}}{\mathit{MoP}_\text{\!max}} \times 100\% \label{eqn:Rsys1c} \\
		& R^\text{sys} = t_3 - t_0  \label{eqn:Rsys1d}
	\end{align}
\end{subequations}
where $R^\text{sys}$ is the system resilience metric; $\mathit{MoP} (t)$ is the measure of performance at time period $t$; and the other variables can be directly found in Fig.~\ref{fig:res-curve}. 

We call this category as the distance-based metrics because their expressions are related to certain distance measurement in a resilience curve, as shown in Fig.~\ref{fig:res-curve}. Equations~(\ref{eqn:Rsys1a})--(\ref{eqn:Rsys1c}) are relevant to the vertical distances while (\ref{eqn:Rsys1d}) is directly a horizontal distance.

These metrics are widely used due to their simplicity.
Reference~\cite{younesi2020assessing} developed several criteria from reliability analysis, e.g. expected demand not supplied (similar as EENS). Extreme value parts were extracted by a filtering mechanism.
EENS was used again in \cite{cai2023resilience} as the resilience metric. Following (\ref{eqn:Rsys1d}), reference~\cite{rosales2019microgrids} applied the time period after a microgrid outage as the resilience evaluation metric.

\subsubsection{Area-Based Metrics}

The second kind is based on the measure of cumulative differences which are expressed as areas in Fig.~\ref{fig:res-curve}. The following are four typical formulations: 
\begin{subequations}
	\begin{align}
		& R^\text{sys} = \int_{t_0}^{t_3} \!\! \left( \mathit{MoP}_\text{\!max} - \mathit{MoP} (t) \right) \mathrm{d} t \label{eqn:Rsys2a} \\
		& R^\text{sys} = \int_{t_0}^{t_1} \!\! \left( \mathit{MoP}_\text{\!max} - \mathit{MoP} (t) \right) \mathrm{d} t \label{eqn:Rsys2b} \\
		& R^\text{sys} = \int_{t_2}^{t_3} \!\! \left( \mathit{MoP}_\text{\!max} - \mathit{MoP} (t) \right) \mathrm{d} t \label{eqn:Rsys2c} \\
		& R^\text{sys} = \int_{t_0}^{t_3} \!\! \left( \mathit{MoP}_\text{\!enh} (t) - \mathit{MoP} (t) \right) \mathrm{d} t \label{eqn:Rsys2d} 
	\end{align}
\end{subequations}
where $\mathit{MoP}_\text{\!enh} (t)$ measures the system performance at time period $t$ when an enhancement strategy is taken. From Fig.~\ref{fig:res-curve}, one can find that (\ref{eqn:Rsys2a})--(\ref{eqn:Rsys2c}) are calculating the areas of $S_\text{M} \cup S_\text{L} \cup S_\text{R}$, $S_\text{L}$, and $S_\text{R}$. But (\ref{eqn:Rsys2d}) instead compares the area between two resilience curves. Note that all these formulas could be normalized to create a ratio version.

These metrics are prevalent options in the literature. This is the case for \cite{gautam2020resilience} as the expected curtailed energy was calculated by (\ref{eqn:Rsys2a}). Beyond the classical results, we further focus on some special extensions below. 
Reference~\cite{senkel2021quantification} found that the resilience curve of heating systems might rebound before returning back to a new normal phase. The authors introduced a deviation bound and calculated the recovery time index by (\ref{eqn:Rsys1d}) and the performance loss index by (\ref{eqn:Rsys2a}), both with additional normalization. 
Reference~\cite{zhao2021resilience} formulated a fluctuating resilience curve and an binary variable was used to select the eligible periods for calculation.

\subsubsection{Angle-Based Metrics}

The third kind exhibits the rate of changes of resilience performance, which can be expressed as the geometric slopes in Fig.~\ref{fig:res-curve}. 
\begin{subequations}
	\begin{align}
		& R^\text{sys} = \frac{\mathit{MoP}_\text{\!max} - \mathit{MoP} (t_1)}{t_1 - t_0} \label{eqn:Rsys3a} \\
		& R^\text{sys} = \frac{\mathit{MoP}_\text{\!max} - \mathit{MoP} (t_2)}{t_3 - t_2} \label{eqn:Rsys3b}
	\end{align}
\end{subequations}

Equation~(\ref{eqn:Rsys3a})(\ref{eqn:Rsys3b}) are called angle-based metrics because their expressions are equivalent to $\tan \theta_\text{L}$ and $\tan \theta_\text{R}$. These two formulas have a clear meaning of the rate of change in resilience performance, which is useful in short term to quantify the performance degradation or recovery speed.

\subsubsection{Statistical Metrics}

The fourth kind involves many statistical indicators. A typical case is shown below:
\begin{align}
	\label{eqn:Rsys4}
	R^\text{sys} = \frac{1}{N(N-1)} \sum_{i \neq j} d_{ij}
\end{align}
where $N$ is the number of nodes and $d_{ij}$ is the shortest path between node $i$ and $j$. This metric is useful for networked systems to quantify the network connectivity. 
A strong connectivity means deep sector coupling and interactions.

Reference~\cite{yodo2021resilience} termed the quantity~(\ref{eqn:Rsys4}) as network efficiency and applied it to analyze a power-gas-petroleum system. The change of network efficiency was captured in each resilience phase under different conditions. A similar sensitivity analysis was conducted in a hydrogen-based multi-energy system~\cite{bartolucci2021hydrogen} to express the resilience index as a function of hydrogen penetration rate. 

\subsubsection{Probabilistic Metrics}

The fifth kind is a broad range of probabilistic indicators that can be extended from (\ref{eqn-dist-mop}) and (\ref{eqn:Rsys4}). Two typical formulations are given: 
\begin{subequations}
	\begin{align}
		\label{eqn:Rsys5}
		& R^\text{sys} = \mathsf{E}_w [R^\text{sys}_w] \\
		& R^\text{sys} = \mathsf{CVaR}_\alpha [R^\text{sys}_w]
	\end{align}
\end{subequations}

Reference~\cite{francis2014metric} applied Monte Carlo integration to calculate the expected system degradation after disruptions. 

\subsection{Comprehensive Evaluation} \label{subsec:comp-eval}

Resilience evaluation of multi-energy systems is a complicated task, so any single metric might not reflect a full picture of the entire system. It is necessary to build up a comprehensive evaluation system using multiple indicators. 

\begin{figure}[t]
	\centering
	\includegraphics[width=0.45\textwidth]{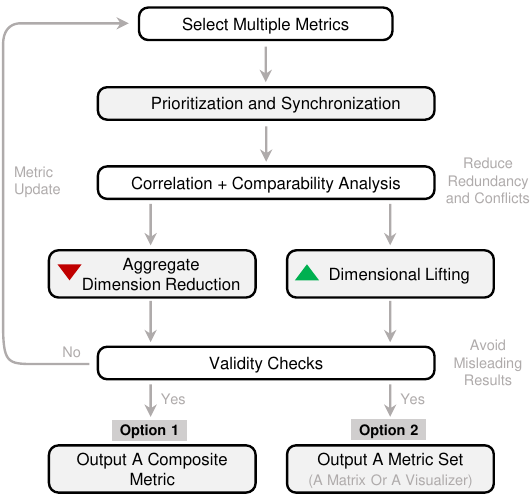}
	\caption{General process of multi-metric evaluation for multi-energy systems. Two popular options are merging into a composite metric or a metric set.}
	\label{fig:eval-flowchart}
\end{figure}

A general process  is established and illustrated in Fig.~\ref{fig:eval-flowchart}. Here, an initial collection of multiple metrics are prepared, and they should be first prioritized and synchronized according to the specific applications. Then, a correlation and a comparative assessment are conducted to reduce the redundancy and conflicts in the results from different metrics. There are two popular options for the next step: aggregate dimension reduction and dimensional lifting. All results must be validated to avoid any misleading results.

\subsubsection{Prioritization and Synchronization}

Prioritization and synchronization are two typical procedures in comprehensive evaluation to handle the interactions across energy sectors.

Prioritization is applied to determine the key energy sectors to evaluate. As extreme events may influence energy sectors differently, the evaluation process should shift the focus to specific sectors that may experience the greatest harm. Prioritization can be realized by ranking or relative importance scoring. In this case, it is possible to enhance the computational efficiency by ignoring minor details.

Synchronization aligns the timelines and resilience curves of different energy sectors (separate curves) and determines a proper simulation time step to evaluate resilience performance. The most common option is to use the event timeline with a proper temporal resolution. Different timescales and sectors will be well synchronized if the temporal resolution is high enough to reflect the disaster and mitigation dynamics. This process is a necessary step to correctly characterize sector coupling, avoid mismatches and conflicts between sectors, and evaluate different sectors in a unified measurement system. There are also options for multi-rate simulation which adopts different temporal resolution (step sizes) and aligns all the sector simulations through time synchronization.

\subsubsection{Composite Metrics}

Aggregate dimension reduction merges diverse sub-metrics into one composite metric. A typical approach is to normalize the sub-metrics and then calculate a weighted sum. This is an extension of the aggregation flowchart in \cite{gasser2020comprehensive} where eight normalization and six aggregation techniques were considered. Reference~\cite{liu2023risk} developed nine risk indices to measure the system operational risk at various levels. Reference~\cite{chanda2018quantifying} normalized the modeling format of different extreme events by using a code-based data structure to record the failure time of different magnitudes (seconds to weeks). Both attempts were useful and informative to handle the heterogeneity of different metrics. 

Reference~\cite{mutani2018energy} developed three indicators to evaluate the multi-energy resilience of residential and tertiary buildings in Turin, and aggregated them into a composite index by a min-max normalization and equally-weighted summation. Reference~\cite{zhang2020resilience} presented the operational and infrastructure indices for electricity and natural gas systems. Indices of different energy systems shared the same format and were summed up to derive the final system index. Another resilience assessment index was established in \cite{zhao2021resilience} to quantify the impacts of multi-energy coordination, while the correlation was handled by an extended analytic hierarchy process method. In \cite{gatto2020measuring}, a Monte-Carlo simulation was conducted to derive different weighting schemes for an interval-based composite metric using the value ranges of individual metrics. Reference~\cite{zhao2021resilienceindex} built an optimal weighting model to aggregate the subjective and objective metrics.

\subsubsection{Metric Sets}

Dimensional lifting maintains all the original measures and concatenates them into a high-dimensional metric set. Normalization can be applied to keep the value ranges close to each other. 

Reference~\cite{moslehi2018sustainability} developed a resilience matrix whose elements were resilience indices in different failure scenarios. Reference~\cite{wang2019integrated} analyzed the role of solar field and carbon capture systems, and the multi-metric evaluation was visualized in a radar chart.

\subsubsection{Practical Considerations}

Evaluation may be affected by a few practical issues. Compound events will influence the prioritization and synchronization steps in comprehensive evaluation. More sectors should be selected and higher temporal resolution is required to capture critical dynamical details. Under compound events, detailed evaluation may increase the requirement of data collection and computational speed.

Also, resilience metric selection heavily depends on the data resources. To illustrate, failure-related metrics require failure rates data; structural metrics require topological data; system-oriented and system-level metrics require sensors frequently capturing the system performance. System-level metrics generally have a higher requirement of data and may suffer from data insufficiency and errors. Comprehensive evaluation is recommended if multiple metrics are available, but the weighting coefficients must be carefully calibrated. 

Note that all the previous discussions belong to quantitative approaches. There are other alternatives in technical reports and industrial projects, e.g. framework analysis, questionnaire designs, and hybrid evaluation~\cite{erker2017resilience}.

\section{Resilience-Oriented Operation and Planning of Multi-Energy Systems} \label{sec:optimization}

This section overviews the canonical forms and recent advances in multi-energy system operation and planning.

\subsection{Canonical Form and Uncertainty} \label{subsec:canoical-form}

Mathematical optimization is widely used in resilience-oriented operation and planning. 
An optimization model is composed of an objective function, several constraints, and an optional uncertainty set. For multi-energy systems, interdisciplinary insights from different energy fields are required.

\subsubsection{Objective Function}

Two common objectives are maximizing the resilience performance or minimizing the total cost. 
There are also special objectives such as regret minimization.

\emph{Maximizing the Resilience Performance}: Most metrics from Section~\ref{sec:evaluation} can serve as an eligible objective function, but the most popular choice is to close the performance gap:
\begin{align}
	\label{eqn:obj-eval}
	\min \ \mathit{MoP}_\text{max} - \mathit{MoP}_\text{min}
\end{align}

Equation~(\ref{eqn:obj-eval}) can measure the load shedding if the two terms are the energy demand in the normal and degradation phase. It is equivalent to maximize the $\mathit{MoP}_\text{min}$ or restore as much energy demand as possible when the original performance $\mathit{MoP}_\text{max}$ is known and fixed. 
Additional weighting factors for different time periods can be added~\cite{li2022coordinating}.

\emph{Minimizing the Total Cost}: A typical cost function is often broken down as follows:
\begin{align}
	\label{eqn:obj-cost}
	\min \ C_\text{inv} + C_\text{opr} + C_\text{other} + \gamma C_\text{pen}
\end{align}
where the four terms $C_\text{inv}$, $C_\text{opr}$, $C_\text{other}$, $C_\text{pen}$ refer to the investment cost, the operation \& maintenance cost, the additional cost (e.g. carbon price), the penalty cost (e.g. constraint violation or emissions); and $\gamma$ is a weighting factor. 

The investment ($C_\text{inv}$) and operation \& maintenance costs ($C_\text{opr}$) have different expressions in the literature, but a typical case is given below~\cite{oh2024bi}:
\begin{subequations}
	\begin{align}
		& C_\text{inv} = \sum\nolimits_l \kappa_l u_l \\
		& C_\text{opr} = \sum\nolimits_i \sum\nolimits_t \left( \kappa_i^\text{var} P_{it} + \kappa_i^\text{fix} u_{it} \right)
	\end{align}
\end{subequations}
where $u_l$ is a binary variable indicating whether to build a new transmission line or pipeline $l$; $\kappa_l$ is a unit price coefficient for installation; $u_{it}$ is another binary variable indicating whether the generator $i$ is online at period $t$; $P_{it}$ is the generator energy output; $\kappa_i^\text{var}$, $\kappa_i^\text{fix}$ are the coefficients for variable (fuel) and fixed costs of energy generation.

Other costs are covered in $C_\text{other}$, e.g. generator start-up costs, maintenance/repair costs, energy storage degradation cost, and thermal loss cost. The penalty cost $C_\text{pen}$ is designed to provide negative incentives for carbon emissions or violating any major constraints. This penalty is often linearly related to the expected energy not served $\mathit{EENS}$.


\emph{Multiple Objectives}: Alternatively, a multi-objective function can balance conflicting objectives by weighting factors. A Pareto frontier further presents multiple trade-off solutions rather than a single optimum.

Reference~\cite{tangi2025designing} formulated four competing objectives including costs, global warming potential, environmental impacts, and reliability. In \cite{zhang2024enhancing}, the cost and resilience were balanced within a bi-objective framework. 
Note that the objective terms for different sectors can be merged into one (weighted sum), which reflects a cross-sector trade-off between the resilience performance of different energy subsystems.

\subsubsection{Constraints}

We next summarize the popular constraints. 

\emph{Energy Flow Constraints}: Any flows must follow the physical laws and the transfer limits (see Subsection~\ref{subsec:evol-mes-model}). In operation and planning, linearized models are generally preferred to ensure computational tractability. These constraints include several energy flow equations and additional inequality constraints such as:
\begin{subequations}
	\begin{align}
		& | \ P_l \ | \le P_l^\text{max}, \quad \forall l \label{subeqn:flow-ublb} \\
		& | \ P_l + \mathit{LODF}_{l \iota} P_\iota \ | \le P_l^\text{max}, \quad \forall l , \forall \iota \neq l \label{subeqn:lodf}
	\end{align}
\end{subequations}
where $P_l$ is the energy flow on the transmission line or pipeline $l$ (this is a concise notation as a placeholder of $P_{ij}$, $G_{ij}$ or $M_{ij}$, see Subsection~\ref{subsec:evol-mes-model}); $P_l^\text{max}$ is the maximum capacity; $\mathit{LODF}_{l \iota}$ is the line outage distribution factor capturing the possible change on line $l$ if another line $\iota$ is out of service. Note that (\ref{subeqn:lodf}) is also termed as a N-1 constraint.

\emph{Radiality Constraints}: Radial topologies of distribution systems must be maintained after natural disasters. 
The minimal conditions of a radial multi-energy network are well-connectivity and no loop.

The connectivity property requires that any energy user is energized by at least one generator. The most straightforward way to enforce connectivity is to apply the energy flow constraints (e.g. (\ref{eqn:dcpf})(\ref{eqn:gas-flow})(\ref{subeqn:flow-ublb})). The connectivity holds if any feasible solution exists. A simplified alternative is the commodity flow model. 

Binary variables $u_l$ are introduced to perform line selection so that zero energy flows are enforced for unselected lines:
\begin{align}
	-P_l^\text{max} u_l \le P_l \le P_l^\text{max} u_l, \quad \forall l
\end{align}

The no-loop property must be satisfied as well. 
During the recovery phase, a multi-energy system can be split into a ``spanning forest'' and continue the energy supply locally. The loop-eliminating approach is popular to remove loops:
\begin{align}
	\sum\nolimits_{l \in \Omega_L} u_l \le | \Omega_L | - 1, \quad \forall \Omega_L
\end{align}
where $\Omega_L$ is a loop in a given graph expressed by a set of line indices; and $| \Omega_L |$ is the set cardinality meaning the number of lines involved in this loop.

A common drawback is that the number of loops may increase rapidly for large graphs, but in practice, real-world systems often have much fewer loops. 
The other options for radiality constraints include path-based models, primal dual graphs, and graph-theoretic approaches~\cite{lei2020radiality}. 

\emph{Operational Feasibility Constraints}: 
This category broadly includes a number of component-level constraints that are problem-specific and greatly vary across different sectors.

A common example is the feasible region model for CHP generators from Subsection~\ref{subsec:evol-mes-model}, and another example is the ramping constraint for gas-fueled generators~\cite{chen2023risk}. 

\emph{Financial Budget Constraints}: Budgets are applied to avoid expensive investments. For system planning, the budgets for transmission lines/pipelines are typically capped by:
\begin{subequations}
	\begin{align}
		& \sum\nolimits_l u_l \le n_\text{max} \\
		& \sum\nolimits_l \kappa_l u_l \le C_\text{max} 
	\end{align}
\end{subequations}
where we abuse $u_l$ to indicate investment selection; $n_\text{max}$ is the maximum number of lines; $\kappa_l$ is the capital cost per line and $C_\text{max}$ is the total limit.


\subsubsection{Uncertainty Sets}

We will next outline the uncertainty sources and the typical formulations of uncertainty sets.

\emph{Uncertainty Sources}: Natural disasters are hard to predict and thus become a primary source of uncertainty. The advanced statistical models from Subsection~\ref{subsec:mod-and-pred} are helpful in nowcasting, but they can hardly guarantee a precise and reliable performance in long term.
Renewable energy is another uncertainty source, because their outputs are influenced by meteorological variables such as solar irradiation or wind speed. 
There are a few other sources such as energy prices, multi-energy consumption, and human behaviors~\cite{li2021stochastic}.

\emph{Uncertainty Set Formulation}: 
There are three popular and representative formulations: scenario sets, box uncertainty sets, and budget uncertainty sets. 

A scenario set refers to a collection of representative situations showing the potential outcomes. 
Monte Carlo simulation is the most popular approach to generate a scenario set from a series of given distributions~\cite{yao2019rolling} (e.g. GEV, log-normal distributions, see Subsection~\ref{subsec:mod-and-pred}). Some studies further conduct scenario reduction after the scenario generation to select the most representative cases. 

A box uncertainty set assumes that the uncertain variables are lying in a hyper-rectangle region. 
A typical box uncertainty set for solar energy output $P_t^\text{solar}$ is expressed below~\cite{li2021stochastic}:
\begin{align}
	\mathcal{U}_\text{box} = \{ 
	P_t^\text{solar}, \forall t \mid \hat{P}_t^\text{min} \le P_t^\text{solar} \le \hat{P}_t^\text{max} \}
\end{align}
where $\hat{P}_t^\text{min}$, $\hat{P}_t^\text{max}$ are the lower and upper bounds, which can be determined by a fluctuation range surrounding a nominal value of solar outputs.

A budget uncertainty set is a polyhedral set that controls the redundancy by a budget coefficient. Below is a budget uncertainty set limiting the number of faulted power lines~\cite{shao2017integrated}:
\begin{align}
	\mathcal{U}_\text{budget} = \{ z_l, \forall l \mid \sum\nolimits_l z_l \le \Gamma, z_l \in \{0,1\} \}
\end{align}
where $z_l$ is a binary variable representing whether line $l$ has tripped; the budget coefficient $\Gamma$ reflects the destructive duration and a larger value generally means more faults.

Note that any uncertainty set combines the variables from different energy sectors is called a coupled uncertainty set. This set reflects the energy sector interactions through a trade-off between different sectors' uncertainty budgets or a underlying mechanism for risk transfer and allocation.

Multiple extensions existed in the literature: ellipsoidal sets, polyhedron sets, distributionally robust uncertainty sets, and decision-dependent uncertainty sets. A less-popular option is the composite sets intersecting two or more simple sets.

%

\subsection{Solution Methodology} \label{subsec:res-solution}

Optimization is still a dominating choice, but the use of evolutionary search and learning-based methods is growing.

\subsubsection{Convexification of Constrained Optimization}

Nonconvex optimization is hard to handle. 
Currently, there are a few general techniques, including piecewise linearization, McCormick inequalities, second-order cone relaxation, and sequential linearization.


Risk-aware optimization typically uses CVaR to capture uncertainty. For these models, 
a specific effort is to convexify the CVaR term as follows~\cite{liu2021resilient}:
\begin{subequations}
	\begin{align}
		\min \quad
		& \mathsf{CVaR} = \mathsf{VaR} + \frac{1}{1 - \beta} \sum\nolimits_w \pi_w \mathsf{SF}_w \label{subeqn:cvar-obj} \\
		\text{s.t.} \ \quad
		& \mathsf{SF}_w \ge x_w - \mathsf{VaR}, \quad \forall w \label{subeqn:cvar-con1} \\
		& \mathsf{SF}_w \ge 0, \quad \forall w \label{subeqn:cvar-con2}
	\end{align}
\end{subequations}
where $\mathsf{CVaR}$ is the conditional value at risk at a probability level of $\beta$; $\mathsf{VaR}$ is the value at risk and $\mathsf{SF}_w$ is the potential shortfalls beyond this value; $\pi_w$ is the probability for scenario $w$; $x_w$ is a placeholder for all possible outcomes. 


\subsubsection{Reformulation of Constrained Optimization}

Robust optimization needs effective reformulation.

Regular robust optimization can be well reformulated by seeking the robust counterparts (worst cases). These robust counterparts convert the uncertainty-aware objectives or constraints into an equivalent deterministic formulation. 
There are quite a few standard results for classical uncertainty sets, including box sets, ellipsoidal sets, polyhedral sets, and all potential joint sets.

Distributionally robust optimization intends to find a feasible solution across a range of distributional shifts, and it is particular useful in case the exact distribution is absent~\cite{zhou2024resilience}. The tractable ambiguity sets are mainly based on discrepancy-based, moment-based, shape-preserving, and kernel-based metrics. 

Two-stage robust optimization is a recent variant where the decision are made in stages (here-and-now, wait-and-see stages). After the first stage, the second-stage decisions are adaptive in response to the uncertainty realization. These models exhibit in the following form:
\begin{subequations}
	\label{eqn:two-stage-ro}
	\begin{align}
		\min_x \quad
		& c^\top x \ + \ \max_d \min_y \ b^\top y - e^\top d \\
		\text{s.t.} \ \quad
		& x \in \Omega_x, \ d \in \mathcal{U}, \ y \in \Omega_y(x,d)
	\end{align}
\end{subequations}
where $x$ and $y$ are the first-stage and second-stage decision variables; $d$ is a uncertain source (variable); $c$, $b$, $e$ are three parametric vectors; $\Omega_x$ is the first-stage feasible region; $\mathcal{U}$ is the uncertainty set; and $\Omega_y(\cdot)$ is the second-stage feasible region as a function of $x$ and $d$. 

\begin{figure}
	\centering
	\includegraphics[width=0.42\textwidth]{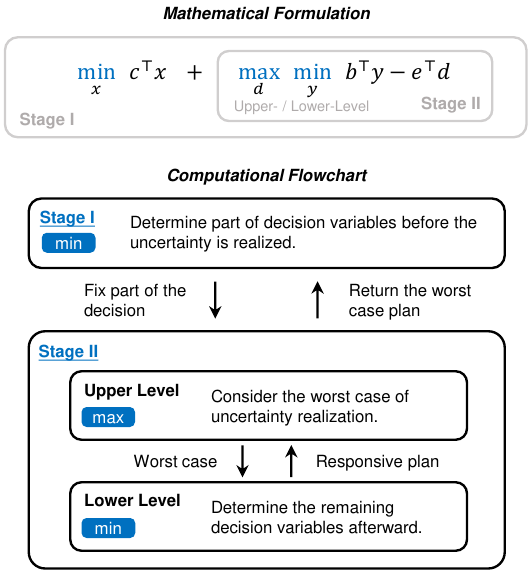}
	\caption{Formulation and computational flowchart for two-stage robust optimization models. These models typically involve two stages, and the intuition is presented in labels as well.}
	\label{fig:two-stage-optim}
\end{figure}

Solving (\ref{eqn:two-stage-ro}) requires dedicated algorithms: Benders-dual cutting plane algorithm and column-and-constraint generation (C\&CG) algorithm~\cite{yan2018coordinated}.
In practice, C\&CG performs more efficient than Benders-dual algorithm (fewer iterations upon convergence). 
Fig.~\ref{fig:two-stage-optim} demonstrates the above computational flowchart of two-stage robust optimizations.

The limitations of robust optimization and its variants include the overly conservative decisions, hard-to-select uncertainty sets, and slow convergence issues. Existing works attempted to improve uncertainty set representation~\cite{xie2025multi} and applied decentralized approach for acceleration~\cite{khojasteh2025distributed}.

\subsubsection{Evolutionary Search}

Evolutionary search is a class of heuristic optimizer inspired by biological evolution and it can find feasible solutions and handle contradicting objectives for highly stochastic and nonconvex optimization models. Typical approaches include the genetic algorithm~(GA), particle swarm optimization~(PSO), and NSGA-II. The concern is on the premature convergence, hard-to-tune parameters, and limited performance guarantees. It is often difficult to determine the proper method prior to multiple trials.

\subsubsection{Learning-Based Approaches}

Learning-based and data-driven approaches have attracted recent interests in the literature, but there is very preliminary effort using these approaches in multi-energy system resilience~\cite{wang2022machine}. This is motivated by the fact that massive historical data could be learned. 
It is expected that machine learning could take advantage of the historical data to make desirable operation and planning schemes. Often, the choice of specific learning models depends on data quality, problem settings, and performance targets. 

\subsubsection{Agent-Based Simulation and Co-Simulation}

Agent-based simulation is a typical computational technique to capture the complex dynamics among decision makers. It accounts for emergent patterns as well as game-theoretic interactions through simulation~\cite{ahmadi2021frameworks}. 
In the literature, some studies were running co-simulation of different energy sectors to determine the optimal operation and planning schemes. A typical strategy is multi-rate simulation which uses different timescales for systems with diverse inertia and applies time synchronization to merge the results. 

\subsubsection{Scalability Consideration}

Among the above choices, learning-based approaches are generally the most scalable option, and they can generate predictions or decisions very fast after training. Constrained optimization models exhibit moderate to poor scalability, especially in the presence of binary decision variables. Their computational burden tends to grow rapidly with problem sizes. In contrast, the scalability of evolutionary search and agent-based simulation are highly problem-dependent, because they support parallel computation but slow convergence may hinder overall performance.

\subsection{Short-Term Resilience-Oriented Operation} \label{subsec:res-operation}

\subsubsection{Operation Scheme}

An operation scheme with resilience consideration is often proposed to enhance the structural (infrastructural) resilience and operational resilience by the coupled/coordinated dispatch of different resources. Structural resilience is essentially the boundary and a necessary condition of operational resilience. 

Structural resilience describes the topological strength of a multi-energy system. The concept of infrastructural robustness was measured in \cite{zuloaga2019resilience} by connectivity, betweenness, demand priority, and demand adjustment metrics for electric and water distribution systems. Co-simulation of the PSLF (for electric grids) and the EPANET (for water systems) software was implemented. Flexible topology has become an important property for structural resilience. Reference~\cite{lin2019combined} carried out the topology reconfiguration and intentional islanding to improve the flexibility of system topology. This flexibility enables the interactions between electric and natural gas subsystems. A similar network reconfiguration strategy was applied in \cite{jiang2020multi} with controllable switches, and a single-commodity flow was utilized to flexibly update the radial topology of distribution networks. Apart from the physical system incidents, there are further discussions regarding communication and sensor failures. Reference~\cite{yang2022resilience} regarded the configuration of PMU and communication facilities as a critical measure to enhance energy resilience against extreme weather disasters. Reference~\cite{jin2022resilience} highlighted the needs to build resilient topologies of sensors and hardware. 

Operational resilience denotes the ability to mitigate performance loss by operational measures, and the temporal-spatial coordination among resilience resources is of great importance. In \cite{ruan2025temperature}, the smart charging of electric vehicles was investigated under cold climates, and the energy management and thermal management were determined by a joint decision model. The temporal coupling of thermal dynamics was applied to mitigate the energy overhead for heating up batteries. Reference~\cite{masrur2022optimized} studied a grid-connected multi-carrier energy microgrid, involving diverse energy sources (such as CHP generators) and demand response (incentivized by time-of-use pricing signals). The temporal shifts of demand and spatial shifts of energy supply provided great flexibility to boost the resilience performance. Reference~\cite{amirioun2018resilience} developed a proactive scheduling scheme and a multi-objective optimization model against hurricanes for multi-energy microgrids. In this system, the energy coupling and interactions came from the gas-fired generators and boilers. In \cite{gargari2023preventive}, another multi-energy microgrids were making use of mobile energy providers and conducting N-1 contingency analysis to determine the vulnerable elements and system risks. A three-stage resilience scheduling approach was proposed in \cite{lv2022multi} integrating multi-level decentralized reserves in a electricity-gas integrated system, and the decision models of each level were solved rollingly. Reference~\cite{haggi2022proactive} established a bilevel model in which the upper and lower level accounted for the operation of distribution and transmission networks. The proposed proactive plan followed a rolling horizon, and long-term backup resources were crucial to provide the desirable temporal shifts for resilience enhancement.

\subsubsection{Operation Under Uncertainty}

Uncertainty is a critical element and extensive efforts have been devoted to capturing and handling the heterogeneous uncertainty factors. 

Stochastic optimization is a common model to take account of uncertainty scenarios. Reference~\cite{yao2019rolling} proposed a rolling restoration model to decide the coupled and coordinated schedule, and the uncertainties in energy demand and branch status were captured by scenario trees through Monte Carlo simulation. Reference~\cite{sun2022resilience} established a stochastic optimization model to minimize the demand curtailment through coordination among different resilience enhancement resources, and a progress hedging algorithm was used to solve the model. Reference~\cite{zakernezhad2021optimal} sequentially solved three single-level stochastic optimization models, while the worst-case scenarios were generated exogenously by setting four-hour facility failures.

A direct extension is to use the two-stage stochastic optimization. Reference~\cite{kavousi2018stochastic} developed a two-stage stochastic post-hurricane recovery model for networked microgrids. A reconfiguration strategy was implemented in the first stage and truck-mounted mobile emergence resources were energized in the second stage. Reference~\cite{li2023restoration} studied a resilient restoration model to coordinate electric and heating systems. A two-stage stochastic program was formulated to handle diverse uncertainties. Another two-stage stochastic model was established in \cite{li2021risk} with the CVaR risk evaluation for multi-energy microgrids. This work considered the voltage/var control, battery degradation, and detailed thermal energy flow. Reference~\cite{li2022enhancing} analyzed how to pre-allocate the trailers and liquefied natural gas to survive an integrated energy system during snowstorms. A two-stage stochastic program with CVaR terms was formulated to capture the resource pre-allocation in the first stage and to minimize the loss of energy demand in the second stage. 
Decomposition strategies are popular to handle the computational burden. These efforts include the progress hedging algorithm~\cite{li2023restoration} and a penalty-based Gauss–Seidel method~\cite{li2022enhancing}.

The two-stage robust optimization has been applied in some recent works. Reference~\cite{dong2024robust} developed a resilience enhancement strategy to get prepared for communication failures in a power-and-thermal cyber-physical system. Decision-dependent uncertainty was formulated and handled by a novel decomposition approach.

\subsection{Long-Term Resilience-Oriented Planning} \label{subsec:res-planning}

\subsubsection{Planning Scheme Structure}

A planning scheme is often developed along with a detailed operational schedule~\cite{yang2025two}. Such a planning-plus-operation structure is widely used in practice because many cost terms can only be measured in a daily or hourly resolution, but these operation details might be simplified for computational acceleration. A planning scheme considered preventive actions and the major mitigation is the investment in new facilities and infrastructural renovation, which greatly improve the structural resilience. 

In the literature, the planning and operation schemes were coupled in \cite{lv2020coordinated} for an integrated electricity-gas energy system. A two-level planning model was formulated by integrating the resilient operation of storage reserve preparation. A possible simplification for the operation part was to merely consider several representative days. Reference~\cite{huang2021resilience} selected three representative solar outputs and energy demand for the summer, intermediate, and winter days. This mitigated the heavy computational burden and the planning model could consider more sector coupling details of generalized energy storage and thermal inertia of heat/cooling loads. Similarly in \cite{ren2023optimal}, three representative days and five typical natural disasters were selected by an extreme natural hazard model and the k-means scenario clustering. These representative scenarios were validated to make the extreme cases visible to planning decisions, and the case study found that the total and marginal cost increasing rapidly with the advancement of system resilience. For seaport multi-energy systems, a planning strategy was proposed in \cite{xie2024fueling} to withstand contingency events, such as extreme weather incidents and equipment failures. The operation impacts were simply considered in the objective function by assuming the contingency scenarios in a whole year. One-year-long operation was a precise option for evaluating the operational cost.

\subsubsection{Planning Under Uncertainty}

Uncertainty makes planning models more complicated and large-scale.

Two-stage robust optimization turns out to be the most popular model for planners because the structure of two stages well align with the planning scheme structure. Reference~\cite{he2018robust} proposed a two-stage tri-level robust optimization model for minimizing the worst-case load shedding of electricity and gas demand. The first level determined the proactive network hardening, the second level identified the potential maximal damages due to natural disasters, and the third level made adaptive scheduling. Reference~\cite{shao2017integrated} presented an integrated planning model to replace line segments of electric grid with underground natural gas pipelines. A two-stage robust optimization was formulated by following the above decision structure, and further discussions on statistical modeling of uncertainty were given as well. Reference~\cite{aldarajee2020coordinated} established a similar tri-level robust model to decide the planning scheme for an integrated electric power and natural gas network coupled via gas-fired generators. The fist level was the planning under normal condition, the second level was the resilience evaluation under extreme conditions, and the third level was the risk-averse re-planning. 

There are quite a lot of different solution approaches for two-stage robust optimization problems. Reference~\cite{zhou2024resilience} developed a stochastic distributionally robust optimization model for the planning of integrated electricity and heat systems, which was then solved by a customized C\&CG algorithm. Reference~\cite{wang2018resilience} transformed the tri-level planning model for an integrated electricity-transportation system into an equivalent bi-level model through optimality conditions and then applied greedy search for the final solution. 
Reference~\cite{li2023robust} studied a resilience-oriented planning for multi-energy network expansion. Additional cut generation was utilized in the second stage to provide better feedback to the planning stage. 
Reference~\cite{saravi2024cooperative} developed the cooperative expansion planning for multi-energy distribution and microgrid systems. Similarly, a two-stage tri-level model was formulated, and it was solved by the combined adaptive dynamic programming and linearized alternating direction method of multipliers~(ADMM).

There are many studies applying stochastic optimization models as an alternative. 
Reference~\cite{javadi2022bi} analyzed the resilience planning flowchart and rolling decisions in three stages. Three stochastic models were formulated to conduct the network hardening, optimal scheduling of energy hubs, and other operational measures. 
Reference~\cite{shao2023risk} established a hydrogen-centered multi-energy microgrid using a risk-constrained stochastic optimization model. The risk constraints were typically enforced by sampling average approximation. 
Reference~\cite{wang2023integrated} presented an integrated planning model under extreme events, which considered cascading effects of system components from a Monte Carlo simulation.

\section{Resilience Pricing and Investment for Multi-Energy Systems} \label{sec:investment}

This section discusses resilience pricing and investment, focusing on investment incentives and trade-offs. 

\subsection{Resilience Pricing Schemes and Incentive Design} \label{subsec:price-incentive}

Resilience pricing establishes price signals for resilience services and planning. Its interdisciplinary nature is challenging because resilience resources are often idle and hard to be consistently valued. Pricing and incentive design must take multiple factors into consideration.

\begin{figure*}[t]
	\centering
	\includegraphics[width=0.9\textwidth]{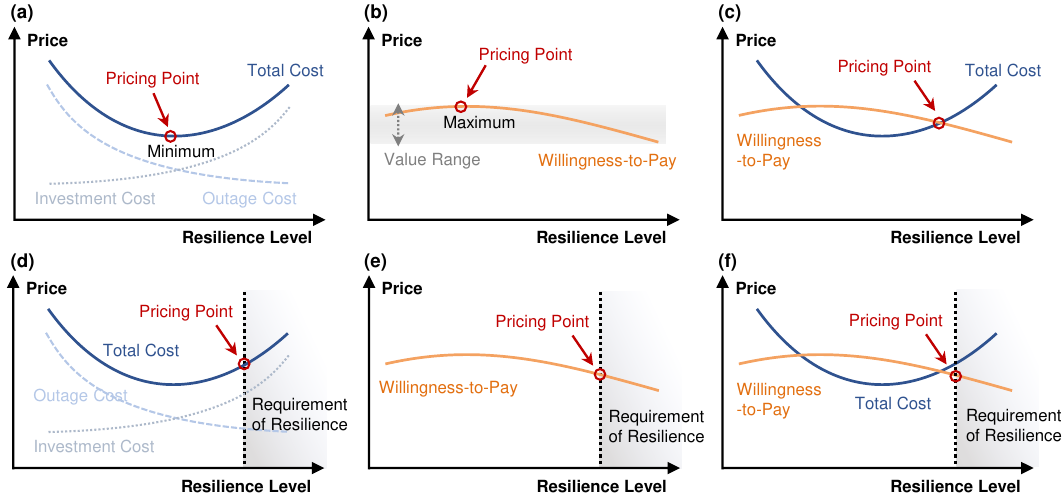}
	\caption{Three typical pricing schemes with and without resilience requirements. The three columns demonstrate the cost-based, value-based, and market-based schemes accordingly, while the two rows compare the conditions without and with the requirements.}
	\label{fig:res-prc}
\end{figure*}

Three typical schemes in the literature include: cost-based, value-based, and market-based pricing.
Fig.~\ref{fig:res-prc} makes a comparison with each column standing for one scheme. We showcase the conditions without and with a required level of resilience performance in the first and second rows.

Next, we will go through these pricing schemes in detail, and then discuss the other influential factors.

\subsubsection{Cost-Based Pricing Scheme}

The price is set according to the total expense. The key is to precisely measure the total cost associated with resilience, which can be further divided into outage cost and investment cost. Each term is assumed to be a function of resilience performance, and any evaluation metrics mentioned in Section~\ref{sec:evaluation} could be used here.

The cost estimation and cost-based pricing schemes (without and with resilience requirements) are given below:
\begin{subequations}
	\begin{align}
		& C_\text{tot} (R) = C_\text{outg} (R) + C_\text{inv} (R) \label{eqn:ctot} \\
		& \mathit{Prc} = \min\nolimits_R C_\text{tot} (R) \label{eqn:ctot-prc1} \\
		& \mathit{Prc} = \min\nolimits_{R \ge R_\text{req}} C_\text{tot} (R) \label{eqn:ctot-prc2}
	\end{align}
\end{subequations}
where $C_\text{tot} (R)$, $C_\text{outg} (R)$, $C_\text{inv} (R)$ are the total, outage, and investment cost with respect to the resilience performance $R$; $\mathit{Prc}$ is the resilience price; and $R_\text{req}$ is the required level of resilience performance. 

Practical applications often add extra profit margin (percentage of revenue) on top of the above benchmark, which is sometimes referred to return on investment (RoI).

Fig.~\ref{fig:res-prc}(a) shows the cost estimates under different levels of resilience. As the performance improves, the outage cost decreases while the investment cost increases at the same time. The total cost appears to drop first, reach the minimum, and then increase rapidly afterwards~\cite{li2016system}. In this case, the pricing point is located at the most cost-effective levels, marked red in the sub-figure. When the minimal levels of resilience are required, the situation shifts to Fig.~\ref{fig:res-prc}(d), and a constraint is enforced in the optimization.

More advances and expansions have been made in the literature. For example, to estimate the investment cost, reference~\cite{lei2022power} proposed a proactive network-constrained economic dispatch model to simulate the resilience performance. This was a resilience-oriented operation model showing how the investment cost might increase with respect to the enhanced resilience. As for the outage cost, a common approach is to estimate the possible load shedding and translate this result into monetary loss by multiplying the VoLL. Reference~\cite{anderson2020integrating} argued that the VoLL might change under different duration of outages. This work parameterized an outage distribution using historical data and applied a dynamic VoLL to reduce the outage cost by 32\% in a large office building in Miami. Reference~\cite{baik2021hybrid} collected a few estimates of resilience cost and found clear variations among the results.
A computable general equilibrium (CGE) model was used to measure the direct economic cost and indirect social cost induced by an outage. 

However, the estimation of outage cost or VoLL is well completed in traditional reliability analysis, but only the impacts of a short outage instead of a long-lasting incident are taken into consideration. A proper monetization of resilience benefits is essential for most of the cost-based schemes.

\subsubsection{Value-Based Pricing Scheme}

The price is set according to the payment willingness of users. Two prevailing concepts are utility functions and willingness-to-pay (WTP). Here, a utility function assigns numerical values to measure the welfare or satisfaction of a user, while the WTP refers to the maximal amount of money that a user would like to spend on a resilience project/service. 

Let $U(R, \delta_\text{inv})$ denote a typical utility function whose inputs include the resilience performance $R$ and a binary indicator $\delta_\text{inv}$ showing whether to request a resilience service or not. Then the WTP and the value-based pricing scheme can be formally expressed as follows:
\begin{subequations}
	\begin{align}
		& \mathit{WTP} (R) = U(R,1) - U(R,0) \label{eqn:wtp} \\
		& \mathit{Prc} = \max\nolimits_R \mathit{WTP} (R) \label{eqn:wtp-prc1} \\
		& \mathit{Prc} = \max\nolimits_{R \ge R_\text{req}} \mathit{WTP} (R) \label{eqn:wtp-prc2}
	\end{align}
\end{subequations}
where $\mathit{WTP} (R)$ is defined as the utility improvement by a resilience service at the resilience level of $R$; and the pricing scheme is to search for the maximal WTP. A typical WTP function is shown in Fig.~\ref{fig:res-prc}(b). 

Estimating WTP is not straightforward in practice because users' preferences are often subjective and application-specific. Questionnaires and proxy approaches are two common options for this task. For instance, an online questionnaire survey of 483 people were conducted in \cite{baik2020estimating}, and the WTP was estimated to be \$1.7--2.3/kWh and \$19--29/day. A proxy approach was used in \cite{moslehi2018sustainability}, and the main idea was evaluating observable actions for resilience purposes, such as the demand-side investment on backup generators.

However, these approaches may yield uncertain and inconsistent results, even for the same group of users. Resilience evaluation is often more challenging than cost estimation.

\subsubsection{Market-Based Pricing Scheme}

Market-based pricing is a combined scheme of the previous two schemes, which relies on the equilibrium to price resilience services. Fig.~\ref{fig:res-prc}(c) visualizes the equilibrium as an intersection of the cost and WTP curves. The equilibrium status can be formally determined using the following formula:
\begin{subequations}
	\label{eqn:mkt-prc}
	\begin{align}
		& C_\text{tot} (R^*) = \mathit{WTP} (R^*)
		\label{eqn:mkt-eq-a} \\
		& \frac{\mathrm{d}}{\mathrm{d} R} C_\text{tot} (R^*) > \frac{\mathrm{d}}{\mathrm{d} R} \mathit{WTP} (R^*) 
		\label{eqn:mkt-eq-b}
	\end{align}
\end{subequations}
where the second equation~(\ref{eqn:mkt-eq-b}) is used to filter out the unstable (invalid) equilibrium.

Given a equilibrium state $R^*$, the pricing schemes without and with resilience requirements are derived as follows:
\begin{subequations}
	\begin{align}
		& \mathit{Prc} = \mathit{WTP} (R^*) = C_\text{tot} (R^*) \label{eqn:mkt-prc1} \\
		& \mathit{Prc} = \mathit{WTP} (\max (R^*, R_\text{req})) \label{eqn:mkt-prc2}
	\end{align}
\end{subequations}

There are efforts to promote this type of pricing scheme. Reference~\cite{clegg2019integrated} developed the nodal pricing for gas demand response to provide extra resilience benefits to an electricity-heat-gas integrated system. In \cite{najafi2020efficient}, a novel market for resilience service was established, where each microgrid submitted bid-quantity energy blocks and the system operator cleared the market during emergency. The equilibrium plan restored unserved loads and avoided the inaccessbility costs of electricity and water services. Reference~\cite{mishra2023resilience} designed a secure transactive energy system with resilience consideration.

\subsubsection{Further Considerations for Pricing}

Resilience pricing in multi-energy systems is shaped by information asymmetry and externalities, which may bias incentive decisions and lead to suboptimal outcomes. Information asymmetry occurs when stakeholders have unbalanced access to relevant information, potentially creating unexpected influences.
%

\begin{figure}[t]
	\centering
	\includegraphics[width=0.42\textwidth]{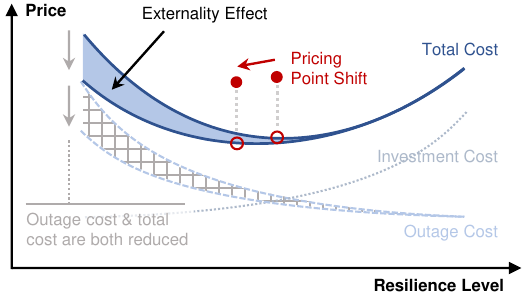}
	\caption{Impacts of positive externality on the resilience cost. Both the outage and total cost are reduced because of the potential support from other energy systems. The pricing point without resilience requirement shifts as well.}
	\label{fig:res-ext}
\end{figure}

Externality effect describes a hidden cost (negative externality) or benefit (positive externality) that is caused by an external stakeholder but is not financially settled in the decision making process. In multi-energy systems, the positive externality may lead to deviation from the optimum, and a widely-used solution is internalizing the externality through subsidies. Fig.~\ref{fig:res-ext} visualizes the impact of positive externality: When the other energy sector is enhanced, the outage and total cost of the target system will drop due to the possible cross-sector support. Taxation, e.g. carbon taxes, is another policy tool with opposing effects (compensate negative external effects). Reference~\cite{schwartz2019utility} demonstrated the positive externality of resilience investment in energy systems by observing that an enhanced electric grid could contribute to the resilience of natural gas, water, or other societal systems. It reported two supporting practices in North Dakota and Wisconsin where multiple sectors coordinated their planning and developed the statewide emergency plans.

\subsection{Resilience Investment} \label{subsec:res-investment}

Resilience investment shares a similar goal of resilience-oriented planning while emphasizing finance, benefit-cost, risk, portfolio, and multi-agent decision-making. It enables cost-effective implementation of resilience plans.

We start with basic categories in the resilience investment. 
The first category comprises federal investment, utility investment, and private investment. 
The second category involves initial investment and maintenance investment. 
The third category includes three types: optimal investment, under-investment, and over-investment. 

The next part focuses on benefit-cost analysis and risk management, followed by interdisciplinary perspectives.

\subsubsection{Benefit-Cost Analysis}

Benefit-cost analysis (BCA) measures the potential monetary benefits/costs of an investment project. BCA uses the Net present value (NPV) to evaluate the cash inflows during an implementation period:
\begin{align}
	\label{eqn:npv}
	\mathit{NPV} = \sum_{t=1}^T \frac{\mathit{CF}_t}{(1 + r)^t} - C_\text{inv}
\end{align}
where $\mathit{CF}_t$ is the future cash flow at period $t$; $r$ is a discount rate; $C_\text{inv}$ is the monetary investment at the beginning. The investment makes profit when $\mathit{NPV}>0$. 

There are two common concepts to define the thresholds from different aspects: internal rate of return (IRR) and break-even time (BET). IRR stands for a discount rate at which NPV of cash flows becomes zero, while BET stands for a time duration over which NPV becomes zero. BET is sometimes called payback period in the literature.

According to the definition, one could calculate IRR and BET by solving the following equations:
\begin{subequations}
	\label{eqn:irr-bet}
	\begin{align}
		& \mathit{NPV} (\mathit{IRR}, T_\text{plan}) = 0 \label{eqn:irr} \\
		& \mathit{NPV} (r_\text{mkt}, \mathit{BET}) = 0 \label{eqn:bet}
	\end{align}
\end{subequations}
where $T_\text{plan}$ is the planned time duration for reaching a balance of investment cost and revenue; and $r_\text{mkt}$ is the average discount rate of the financial market. Using these concepts, an investment becomes profitable when $\mathit{IRR} > r_\text{mkt}$ or $\mathit{BET} < T_\text{plan}$.

Estimating cost and value is a critical topic in BCA. Reference~\cite{broderick2021performance} made efforts in collecting different kinds of cost and benefit from the perspectives of diverse stakeholders. Reference~\cite{esteban2014post} analyzed the resilience investment after the 2011 Fukushima nuclear accident and found that more renewable penetration could achieve better resilience performance. The investment in solar energy was expected to become financially profitable after around 14 years.

\subsubsection{Risk Management}

Risk management is a systematic process to measure and manage risks using proactive portfolios. This is important for multi-energy systems because uncertainty might harm the investment revenue.

Risk preference is an important concept in risk management, describing the attitude towards taking or avoiding risky outcomes. A general believe is that investors are often risk-averse rather than risk-taking in long-term resilience investment, which leads to a under-investment issue. Reference~\cite{mays2022private} studied this problem in generator winterization during the Texas blackout crisis in 2021. It was found that decentralized markets were often prone to under-investment in resilience because incomplete risk trading might disincentive the risk-adverse investors. Mandatory contracting obligations or a capacity market could be two remedy solutions.

\begin{figure}[t]
	\centering
	\includegraphics[width=0.42\textwidth]{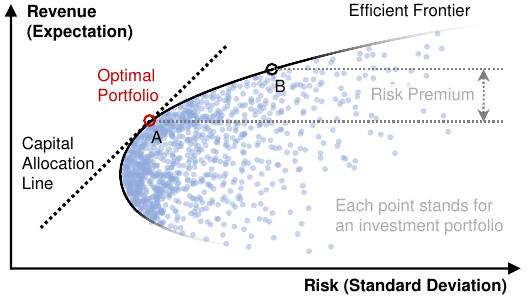}
	\caption{Illustration of investment portfolio management. Different portfolios are expressed as a point while the efficient frontier represents those portfolios that maximize the revenue under different specific risk levels. Risk premium is also demonstrated using two points on the efficient frontier.}
	\label{fig:res-frt}
\end{figure}

Financial products hedge risks and accommodate different risk preferences. The portfolio management is used to identify preferred combinations of resilience products or services.
This process is visualized in Fig.~\ref{fig:res-frt}. An efficient frontier is generated by identifying those revenue-maximizing portfolios under different risk levels. Once can set up the capital allocation line based on the risk preference and find the optimal portfolio (point A). We further highlight the risk premium between the point A and B, which is the additional monetary compensation required for taking a larger risk level.

Reference~\cite{wyrwoll2022reforming} introduced the benefits and issues of sustainability-linked bounds. Reference~\cite{leisen2019regulatory} stated that the design of any financial products should align with the policy regulation to avoid the regulatory risks.

\subsubsection{Interdisciplinary Perspectives}

Resilience investment is a typical interdisciplinary topic that consists of the system operation (physics), the financial settlement (finance), and the human decisions (society/psychology). It is clear that cross-domain knowledge and understanding are important.

Reference~\cite{valinejad2021multi} formulated a multi-agent model to integrate social behaviors and computational social science in power system resilience. Similar agent-based approaches were also used in \cite{gao2016network} to model and predict the human irrational behaviors in the disturbance of extreme events.

\section{Practical Implementation} \label{sec:project}


\subsection{Europe} \label{subsec:proj-europe}

The exploration and application of multi-energy systems took place early in Europe with a rich collection of representative projects. The goal of many European projects are mainly on the coupling of different energy infrastructures and the growing dependence due to demand-side electrification.

The Strategic Energy Technology Plan~\cite{euro2020set} was launched by the European Commission to enhance the resilience and security of energy infrastructure (e.g. electricity, gas, heat \& cold, transport). In November 2018, the working group proposed an implementation plan for energy systems with special attention to demand-side electrification, energy market, integrated energy infrastructure, and digital support. The implementation plan also emphasized the importance of coordinating integrated energy systems and optimizing the use of CHP generators, geothermal and hydrogen resources.

The Center on Regulation in Europe (CERRE) conducted a recent research project in 2023, focusing on the energy resilience of European electricity and natural gas infrastructures~\cite{baldursson2023building}. The project team highlighted the need of new interdisciplinary resilience metrics to take account of the growing dependence and feedback between the electricity and natural gas sectors. This project conducted six case studies to validate how extreme weather and geopolitical factors could threaten energy resilience.
It is still a challenge to evaluate the risks propagating via sector coupling, and extended discussions were given regarding the energy dependence impacts and multi-energy coordination potential in Europe.

The U.K is currently leading the resilience projects of multi-energy systems in Europe. One commonly use concept by the U.K. community is the whole-energy systems, which shares a similar meaning of multi-energy systems.

The Integrated Development of Low-Carbon Energy Systems (IDLES) project~\cite{epsrc2018details} was implemented in the U.K. and funded by the Engineering and Physical Sciences Research Council (EPSRC) for 2018--2023. This project conducted a whole-energy systems analysis to coordinate the complex interactions within the energy domain and identified the role of offshore wind and nuclear energy in compliance with the 6th carbon budget.
The package four of IDLES centered on the resilience and risk management of future whole-energy systems, and multi-energy microgrids were expected to play a fundamental role in securing the decarbonized energy systems in the future~\cite{zhang2020multienergy}. Advanced control of multi-energy microgrids at the local district level was claimed to be more cost-effective than the asset redundancy at the national level.


The Multi-Energy Hubs for Green Mobility project was completed by the Energy Networks Association (ENA) in 2022~\cite{ena2022resilient}. The funding agent was the U.K. Energy Research Center~(UKERC). This project focused on the rail decarbonization using a coordinating and flexible energy supply. The project team investigated the feasibility of multi-energy hubs around 2500 railway stations to address the challenges of energy efficiency, emission reduction, uncertainty and risk management. Resilience strategies of vehicle-to-grid, vehicle-to-station, vehicle-to-train, demand response, and frequency/voltage support were all crucial in the overall resilience performance. It was concluded that rail decarbonization require network upgrade of over \pounds 1 million/km.

The Residential Whole System Integrated Resilience (REWIRE) project~\cite{ena2023rewire} and the Whole Energy System Resilience Vulnerability Assessment (WELLNESS) project~\cite{ena2023whole} were another two ENA projects funded through the U.K. strategic innovation fund in mid-2023. 

The REWIRE project~\cite{ena2023rewire} validated the technical viability and economic benefit of local multi-energy storage facilities. This project improved energy security and resilience to consumers using the potential of power-to-gas, gas-to-power, and local hydrogen storage technologies. It claimed that over 2~TWh wind energy was avoidable nationwide and consumer bills could be greatly reduced. The solution also used low-pressure natural gas pipelines as a transport pathway for hydrogen or as local energy storage. 

The WELLNESS project~\cite{ena2023whole} provided empirical evidences and a novel approach to developing resilience standards for whole energy systems. A resilience assessment model was developed to capture multi-energy flexibility and identify the extreme conditions on the infrastructures. One key task was to investigate the effective applications of flexibility resources including EVs, mobile generators, and multi-energy assets. The cost-benefit analysis and business use cases were conducted to inform the system designs and demonstrations.

In Germany, the Resilience of Integrated Energy Networks with a High Share of Renewable Energies (ResiliEntEE) project~\cite{projectteam2017resilientee} was a research project led by the Hamburg University of Technology during 2017--2021. This project explored the technical and economic solutions to increase the energy resilience through diverse conversion and storage technologies, including power-to-gas, power-to-heat, CHP, and fuel cells. An open-access research toolkit TransiEnt was refined to perform dynamic simulations of integrated energy systems and analyze the carbon emissions and operational costs. A simulation testbed for the Humburg energy systems was developed and empirically validated.

In Switzerland, the Sustainable and Resilient Energy for Switzerland (SURE) project~\cite{sure2024sustainable} was funded by the Swiss Federal Office of Energy for 2021--2027. This project addressed the need for evaluating the sustainability and resilience for a deep decarbonized energy system. The team developed a model called FlexECO for interdisciplinary decision making at all levels. In addition, a case study was conducted to analyze the transport and chemical industries. It was found that spatially distributed grid connections could reduce peak loading on Swissgrid by up to 27\%. Several decarbonization pathways for different sectors such as chemicals, pharmaceuticals, and food production were jointly optimized.

\subsection{The United States} \label{subsec:proj-na}

The U.S. has experienced several billion-dollar natural disasters every year, and great attention is given to the energy resilience. It is found that the U.S. recent practices of multi-energy systems and energy resilience have demonstrated a greater diversity.

Wildfires frequently occur in California during the summer months. Reference~\cite{moreno2022microgrids} discussed the mitigation strategy of using distributed energy resources in microgrids, and the recovery was found significantly ameliorated. A typical example was the Blue Lake Rancheria microgrid (with 500~kW solar) implemented by the Blue Lake Tribe and the energy research center of Humboldt State University. This demonstration microgrid reduced the wildfire risks using CHPs, solar, and storage energy as part of the mitigation actions against natural disasters. In September 2018, the California Public Utilities Commission made the first order of business establishing exemplar microgrids that could enhance resilience performances. The PG\&E company was encouraged to promote community microgrid programs as well as several projects of mobile substation generators.

Publicly shared in \cite{database2023microgrid}, there are 110 microgrid sites locating in California with more than 462~MW total capacity. These microgrids covered a broad range of use cases such as city/community, hospital/healthcare, agriculture, airport, and commercial buildings. As for resilience strategies, many projects applied the multi-energy solutions such as CHPs, fuel cells, and natural gas or diesel backup generators. The implementation of such solutions was likely to increase the dependence across different energy sectors, and cross-disciplinary teamwork is essential for success.

The Hunts Point Resilience project~\cite{nycedc2020hunts} was a consultant project launched during 2016--2019 by the New York City Economic Development Corporation. This project made rigorous assessment on the climate risks and vulnerability, and conducted a benefit-cost analysis for various resilience strategies. The project team found that a flood-resilient tri-generation facility (within 5.2-MW microgrid) could protect the food-related business in the Hunts Point during emergency, and this facility was able to provide heating, cooling, and electricity in a coordinated fashion. 1.1~MW mobile diesel generators were validated to provide energy supply to critical business services in Hunts Point for over three days.

Energy resilience is given high priority in disaster preparedness for local governments. Reference~\cite{cnt2022integration} was a white paper published by the North Central Texas Council of Governments to introduce the best management practices of different actions local governments could take to enhance energy resilience. Among all the mitigation actions, CHP was highlighted as a cost-efficient and popular solution for emergence response. This was demonstrated through a case study of a 50-MW microgrid CHP system at the University of Texas at Austin. Local governments were suggested to incentivize stakeholders by subsidizing the capital investment of CHPs.

In addition to the above projects, the U.S. national laboratories play a leading role in promoting energy resilience. 

National Renewable Energy Lab (NREL) provided a wide range of integrated energy solutions to establish resilient energy systems~\cite{nrel2024resilient}. NREL offered technology and market advice to support interdisciplinary resilience decisions. In particular, NREL supported energy planning decision with the integration of other services such as water, transport, and emergence response. Their focus was also on the use of microgrids, solar plus storage, energy-water coupled systems.

Pacific Northwest National Lab (PNNL) developed the integrated water power resilience project~\cite{pnnl2020integrated} in 2020 to enhance water and power resilience through coordinated planning and operation. Over 30 water and power system industries were interviewed. 
The project team simulated three major U.S. electric grids to demonstrate the benefits in long-term economic viability and risk management. Comprehensive Sankey diagram analysis was recorded as a key project deliverable.

\subsection{Asia} \label{subsec:proj-asia}

Asia countries have made progress in the preparing resilience resources for multi-energy systems. Several typical practices are selected and described in detail.

Singapore-ETH centre was established in 2010 to promote global environmental sustainability, and one of the joint efforts was the Future Resilient Systems project~\cite{frs2024future}. Following a system-of-systems framework, this project had evaluated the system interdependence and resilience performance of a coupled transportation, power, and economy systems.

China's wide geographical regions are prone to many natural disasters. The 2008 great winter storm in South China was one well-known example~\cite{yan2019coordinating}. During this storm, 800 transmission tower collapses and 400 thousand line disconnections occurred in a short period and soon triggered a severe cascading failure. Despite the advanced monitoring and de-icing techniques (e.g. automatic identification of ice thickness), the system operators applied mobile de-icing devices (a vehicle-mounted facility) and conducted coordinated scheduling of electric--transportation systems for emergency preparedness and operation. Efficient traffic prediction, control and routing schemes played a special role in emergency supplies.

Southeast Asia has many remote communities without reliable and affordable energy access. Rural microgrids and distributed energy resources provide a viable solution. The Association of Southeast Asian Nations~(ASEAN) Power Grid Financing initiative~\cite{aris2020asean} is a flagship effort led by ASEAN and coordinated by the World Bank and Asian Development Bank to expand transmission networks to accelerate renewable energy transfer across Southeast Asia. It has successfully united governments, utilities, local institutions, and manufacturing companies. Typical outcomes of the first-phase implementation included 300~MW energy capacity trade between Malaysia, Thailand and Laos; 600~MW electricity export from western Indonesia to Malaysia. This initiative remains active and is advancing the next-phase implementation.

\section{Emerging Challenges and Opportunities} \label{sec:challenge}

\begin{figure}
	\centering
	\includegraphics[width=0.45\textwidth]{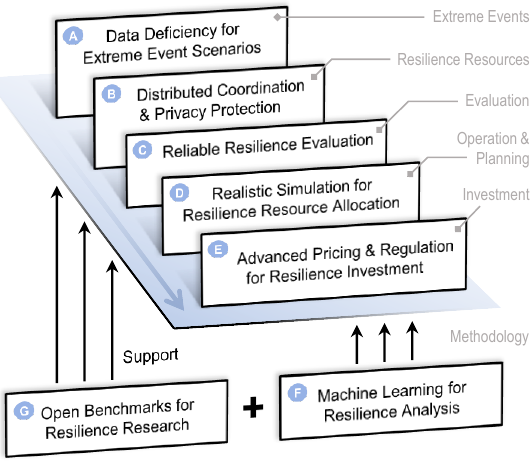}
	\caption{Summary of the emerging challenges and future opportunities. Five research phases and two methodological developments are covered.}
	\label{fig:challenge}
\end{figure}

This section summarizes seven emerging challenges and opportunities for future research, as shown in Fig.~\ref{fig:challenge}.

\subsection{Data Deficiency for Extreme Event Scenarios} \label{subsec:dist-shift-ext}

\subsubsection{Insufficient Data Resources}

The discussion in Subsection~\ref{subsec:dist-shift} has highlighted the distribution shift issue. It can significantly reduce the relevance of historical datasets and cause significant biases in trend extrapolation. Under distribution shifts, reliable data resources might be very limited, especially when our focus is on the distributional tails. Pure data-driven models (e.g. deep learning) suffer more in this settings than physics-based models, so it is recommended to enhance data-driven models with interdisciplinary knowledge (statistics and climate science) for better boundary conditions and more efficient formulations.

Data issues under distribution shifts can be addressed by: correction, generation, and efficient designs.

Data correction follows a simple idea to debias the shifted distributions. This directs toward the observability and predictability of distribution shifts, and the goal is to approximate how distributions are changing over time. Also, distribution shifts may have diverse impacts on probability quantiles, which are often problem-specific. 

Data generation extends the dataset size with simulation data, and high-fidelity simulators can be utilized to generate the expected amount of data and integrate climate fluctuations into the generation process. Many climate models are eligible to provide extra data, and one representative example is the reanalysis data. Reference~\cite{qu2025prolonged} relied on the data from ERA5 and MERRA2 to discover prolonged wind droughts in a warming climate, which exhibited a hidden risks of 20\% reduction of wind power generation. Reference~\cite{ding2024reanalysis} applied reanalysis and ground station data to enhance wind power forecast accuracy. A large-scale extreme weather dataset was established in \cite{kim2025large} which contains global annotations for atmospheric rivers, tropical cyclones, and atmospheric blocking events from ERA5.

Data-efficient designs aim to reduce the data volume required for fitting a forecasting model. A data-efficient model must be able to efficiently learn without dependence on vast datasets. Many existing models cannot work well in this data insufficiency setting. It was found in \cite{liang2023uncertainty} that deep learning algorithms were fragile and lacking robustness in capturing building energy use with distribution shifts. Similar limitations of learning-based weather forecasting models (pure data-driven) were discussed in \cite{bonavita2024some}, and the finding was that these models might not exhibit the same fidelity and physical consistency as physics-based models. In this sense, hybrid models (model-guided plus data-driven) could be a good option because they made use of data resources and domain knowledge simultaneously~\cite{ruan2022improving}.

Another popular choice is few-shot learning, which applies domain adaptation and transfer learning. It can adapt and learn new experience obtained from similar tasks. Reference~\cite{kong2025enhanced} developed an unsupervised few-shot learning approach for landslide susceptibility mapping. Results showed that meta-learning performed better than classical learning-based approaches in areas with limited landslide data.


\subsubsection{Robust Decision Making}

Due to distribution shifts and data deficiency, it becomes challenging to make effective decisions in multi-energy systems. Reference~\cite{fu2017integrated} assessed the resilience of the Great Britain grids in an evolving energy landscape and found only 5--10\% increase in the intensity and frequency of wind storms could fail the system. Reference~\cite{bennett2021extending} validated that climate-induced hurricanes posed great risks to energy systems in Puerto Rico and the climate change impacts were projected to cause a 50\% cost increase by 2040. Since the resource adequacy and resilience were likely to degrade due to distribution shifts, system operators must fully understand the impacts of distribution shifts and take proactive measures to address the negative consequences. 

A typical solution is to continually update prediction or decision models with new data, which can mitigate bias as distributions gradually converges. Reference~\cite{kholodovsky2021generalized} conducted spatio-temporal threshold clustering to automatically identify the distribution shifts in precipitation datasets. Another solution is to use distribution-free optimization and novel risk metrics for reliable decision-making.



\subsection{Distributed Coordination and Privacy Protection} \label{subsec:dist-coord-privacy}

\subsubsection{Distributed and Hierarchical Management}

Multi-energy systems require decentralized coordination among operators. Two mainstream paradigms are hierarchical and peer-to-peer optimization, and both of them can protect data privacy by controlling data access and applying encryption or differential privacy techniques.

%
%

For hierarchical optimization, a typical system includes a coordinator and several independent agents. These agents will submit their individual decisions to the coordinator who will decide the coordinating signal to incentive resubmissions. This entire process is iterating until convergence. Reference~\cite{wang2021resilience} applied hierarchical optimization in an electricity-transportation system where a coordinator determined the energy transfer between different microgrids and each microgrid made the self-dispatching. 

Classical computational approaches for hierarchical optimization involve game theory~\cite{li2023distributed}, dual decomposition, ADMM, and heuristic algorithms. Note that there are similar coordination models in distributed control, which offer valuable insights and knowledge as well. Reference~\cite{li2023distributed} discussed the energy trading for networked multi-energy microgrids, and a tri-layer Cournot Nash game was formulated to capture the heterogeneous uncertainty factors. Reference~\cite{casagrande2022resilient} employed distributed model-predictive control to operate a group of microgrids and restricted the local information sharing for privacy protection.


\subsubsection{Peer-to-Peer (P2P) Management}

Peer-to-peer management typically refers to a distributed architecture that each agent is allowed to equally connect and operate. Without any coordinator, agents will make and update their individual decisions according to the neighborhood until final convergence is achieved. Here, the most classical P2P algorithm is the consensus algorithm. Reference~\cite{rezaei2022stochastic} applied P2P trading between energy hubs during windstorms and survived more energy users. A variety of microgrid studies were utilizing P2P management in post-disaster resilience recovery and pre-disaster resilience enhancement~\cite{spiliopoulos2022peer}.

The absence of coordinators in P2P management enhances privacy protection compared to hierarchical approaches. Ongoing efforts focused on developing advanced algorithms for P2P privacy protection. Reference~\cite{tanis2025cooperative} considered a peer-to-peer marketplace that allowed direct energy trading among prosumers. Reference~\cite{veerasamy2024blockchain} further adopted blockchain-enabled double auction mechanism for P2P trading, and maintain resilience against false data injections by federated learning and local differential privacy.


\subsection{Reliable Resilience Evaluation} \label{subsec:rel-res-eval}

Resilience evaluation is more complicated for multi-energy systems than any single system because of diverse metrics, multiple stakeholders, and high technical complexity. In this context, data quality, metric selection, and method effectiveness may all become bottlenecks. Performing reliable evaluation has become more essential than ever before. Reference~\cite{hossain2021metrics} pointed out three data issues of natural disaster events: small amount, limited accuracy, and difficult-to-measure biases. Reference~\cite{pandey2020resiliency} would never achieve promising outcomes in distribution system reconfiguration without data from costly distribution PMUs.

Most evaluation metrics rely on well-defined resilience curves, but according to \cite{nichelle2021extracting}, characterizing these curves by real-world utility data could be challenging, and even separating the accident and recovery phases might not be intuitive. This highlights a urgent need to conduct more realistic and reliable explorations for evaluation metrics.

Reference~\cite{espinoza2020risk} repeated simulations to evaluate the role of infrastructure components against earthquakes by a (conditional) value-at-risk metric. Reference~\cite{bollinger2016evaluating} developed a similar approach to repeatedly assess the resilience against floods and heat waves. A sensitivity analysis was conducted to identify the criticality-adjusted vulnerability and improve the reliability of final results. 

Currently, there is no mature industry standard/guidance for resilience evaluation of multi-energy systems. Reference~\cite{ieee2020guide} is an IEEE guidance for distribution network resilience that provides the definition and covers how to measure and report resilience metrics. Other relevant material is an IEEE recommended practice~\cite{ieee2014recommended} for mitigating disaster impacts in petroleum and chemical industry. 


\subsection{Realistic Simulation for Resilience Resource Allocation} \label{subsec:real-simul}

\subsubsection{High-Fidelity Models}

Since extreme events are rare, the system simulation plays a special role in resilience studies and current scenario analysis often rely on the extensive use of simulation data. Establishing a high-fidelity simulator can contribute to more informative and trustworthy scenarios that might be ignored in coarse-grained simulation.  

Reference~\cite{clegg2019integrated} developed a high-resolution spatial-temporal model to evaluate the low-carbon heating options in an integrated electricity-heat-gas system. Reference~\cite{ascione2017resilience} used the EnergyPlus software to formulate a high-resolution building energy model for evaluating the climate resilience.  Reference~\cite{confrey2019energy} implemented different types of grid fault patterns in OpenDSS and generated solar panel outputs by a system advisor model developed by NREL. Reference~\cite{nasiri2024decentralized} discussed the cold load pickup phenomena, describing an overcurrent condition when a distribution network was re-energized after a prolonged outage. This issue could be clearly detected and addressed through high-fidelity simulation.

There are progress in high-fidelity simulation, but most works do not take full considerations of resilience issues, e.g., they rarely consider the coupled disruptions or cascading failures across energy sectors. In this context, their adaptability to resilience use cases might be overestimated.

\subsubsection{Computational Acceleration}

Operating a multi-energy system is inherently complex. The computational burden may further increase when a variety of extreme events or high-fidelity component models are used. It is urgent to design simulators and acceleration algorithms.

In industrial projects, decomposition and simplified modeling have been widely utilized to reduce the computational complexity. Reference~\cite{ruan2025temperature} proposed a reduced-order dual decomposition method to solve a large-scale charging decision problem, and this decomposition turned out to achieve a better solution within a tight execution time limit.  Reference~\cite{ding2020multiperiod} used an auxiliary inducing function to accelerate the computation of mixed-integer linear program. Similarly, a dual decomposition algorithm was given in \cite{ma2019resilience} to accelerate the resilience-oriented design of distribution systems.


\subsection{Advanced Pricing and Regulation for Resilience Investment} \label{subsec:scar-prc-inv}

\subsubsection{Scarcity Pricing}

Effective pricing of resilience resources is the key to investment in multi-energy systems. However, the classical marginal pricing cannot compensate the huge risk premium of resilience resources. An alternative is the scarcity pricing, which allows a price above the marginal cost when the supply-demand balance becomes tight.

There is a lack of studies discussing scarcity pricing in multi-energy resilience, but the practical knowledge derived from electricity market (capacity and reserve markets) can provide valuable insights. For example, the scarcity pricing options in European balancing markets were introduced in \cite{papavasiliou2021market} to discover the underestimated value of real-time reserve capacity. Reference~\cite{bajo2021operating} focused on the ERCOT market in Texas, where the design of operating reserves were required to upgrade in response to the increased renewable energy penetration. A recent efforts in \cite{fabra2022learning} pointed out that private incentives were insufficient for adverse extreme events, while electricity capacity mechanisms provided valuable lessons to incentive the reserve auctions.


\subsubsection{Public Regulation}

Public regulation is often necessary in trading scarce resources to prevent market failures or undue market power. Currently, there is little literature on investment regulation of multi-energy systems. 

The regulation progress for blackout events is informative to deepen a general understanding. In particular, a thorough discussion was conducted in \cite{martin2021power} on how to design financial incentives for renewable energy and storage in Australia, and it found that 24\% of stakeholders required policy support for climate and storage development. 
Reference~\cite{mummery2025rethinking} focused on the network regulation for climate resilience, and pointed out a poor coherency between Australia's energy policy and resilience requirements. Reference~\cite{rahman2025unveiling} evaluated how global policy and regulatory frameworks might promote renewable energy transitions for climate resilience. This might not be surprising as the discussions in this area were rare. 


\subsection{Machine Learning for Resilience Analysis} \label{subsec:dl-res-anls}

\subsubsection{Reinforcement Learning}

The rapid advances in machine learning and computing hardware open an interdisciplinary opportunity for resilience research in multi-energy systems. Machine learning has great potential in every resilience phase: upgrade the extreme event estimators, explore the hidden resilience resources, boost the computation of operation \& planning, and improve the resilience pricing schemes.

In resilience studies, the use of machine learning can be categorized into: reinforcement learning~(RL) and other learning approaches. It is well known that RL is a sequential decision-making approach that rewards desired behaviors and punishes undesired ones. RL heavily depends on reliable environment simulator in the Markov decision process. Reference~\cite{nik2025adaptive} incorporated adaptive reinforcement learning into the energy management process to boost climate resilience and building energy flexibility in 17 future climate scenarios for 2040--2069. Reference~\cite{wang2022multi} formulated a partially observable Markov game as the environment simulator to train a multi-agent RL model for mobile energy storage systems.

New techniques have been developed to improve RL performance. 
Reference~\cite{qiu2023hierarchical} formulated repair crew scheduling as a decentralized partially observable Markov decision process and applied hierarchical multi-agent RL with proximal policy optimization.
Reference~\cite{zhang2023bayesian} developed a Bayesian RL model for multi-energy microgrids, which can achieve near-optimal policies and stable training.


\subsubsection{Other Learning Approaches}

A variety of machine learning models have been applied. 
Reference~\cite{yang2024comparative} proposed a novel model for load forecasting under extreme wildfires. A wavelet-based machine learning model was developed in \cite{yeditha2020forecasting} to estimate the extreme flood events in flood-prone basins using the satellite precipitation data. A hybrid model combining neural networks and optimization was developed in \cite{ruan2020neural} to accelerate multi-agent coordination. 

Machine learning exhibits competence in simulating future scenarios of extreme events. Reference~\cite{kurth2023fourcastnet} formulated a data-driven Earth system simulator adapting Fourier neural operator for NWP acceleration. 
Reference~\cite{ding2019modeling} introduced an extreme-value loss to detect long-tailed distributions.
Pareto loss and Kurtosis loss were found useful to improve the prediction of heavy tails. 

In addition to forecasting, machine learning is capable of establishing resilience metrics and generating extreme event scenarios. Reference~\cite{utkarsh2022self} developed a self-organizing map model to calculate the time-varying resilience metric, which was then embedded into a resilience resource allocation system as feedback. In \cite{he2025systematic}, the physics-informed neural network was applied for resilience assessment. Reference~\cite{klemmer2021generative} utilized a generative-adversarial network to create spatial-temporal weather patterns to analyze extreme weather events.


\subsection{Open Benchmarks for Resilience Research} \label{subsec:benchmark}

Research on multi-energy system resilience faces a common challenge in reproductivity and insufficient public resources. It is usually hard to compare different approaches fairly because many resilience studies are utilizing diverse data sources or simulators. This is partially attributed to the methodological diversity and limited attention from the community.

Efforts in this area remains limited. North American Energy Resilience Model (NAERM)~\cite{staid2021north} is developed by Sandia National Laboratory to simulate the energy infrastructures and interdependent systems. Micropolis~\cite{francis2011probabilistic} is a virtual city of 5,000 residents launched by Texas A\&M University as a testbed for infrastructure risk research, and the synthetic power and water distribution networks are established with full details. Other national labs and universities have managed similar tools but the current level of software functionality and openness is clearly insufficient. 

There are some pioneering simulators in power system resilience area. Reference~\cite{murphy2020adapting} studied resilience simulation and evaluation by adapting five existing models which were not originally built for resilience. Reference~\cite{wu2021open} developed an open-source extendable model to realistically analyze the 2021 Texas power outage. The key network topology was generated by a technique called synthetic networks~\cite{birchfield2016grid}.

It is imperative for the global energy community to advance public data and platforms. International organizations such as IEEE, IET, and CIGRE can organize subcommittees or task force to accelerate the progress in this domain. 

\section{Concluding Remark} \label{sec:concl}

Enhancing climate resilience of energy systems is essential for reliable supply, and a promising solution is to facilitate cross-sector coordination and resource sharing.
%

This review is focused on how a multi-energy system can exploit resilience resources against climate-induced extreme events. We summarize the latest progress in extreme event modeling (Section~\ref{sec:extreme-event}), resilience resources (Section~\ref{sec:resource}), resilience evaluation (Section~\ref{sec:evaluation}), resilience-oriented operation \& planning (Section~\ref{sec:optimization}), resilience pricing \& investment (Section~\ref{sec:investment}), and practical implementations (Section~\ref{sec:project}). This paper further showcases the future challenges and opportunities (Section~\ref{sec:challenge}).

The major insights of this work are collected below:
\begin{enumerate}
\item Nested coupling (sector coupling and coupled disruptions) is the key challenge of resilient multi-energy systems. Further efforts are required to analyze coupled disruptions such as compound events.

\item Distribution shift offers a lens to track climate change impacts in terms of intensified extreme events and drifting mean values. Future work must address the uncertainty in identifying and estimating these shifts.

\item Data deficiency poses a challenge to estimate extreme events and make robust decisions. Future solutions should merge climate simulators, data-driven models, and advanced optimization.

\item Distributed coordination across energy sectors has not been well investigated. Future work must consider the inherent heterogeneity issues.

\item Scarcity pricing is a promising solution to assess the value of resilience resources. Future work can dive into the investment structures and methods.

\item Machine learning does not reach its full potential in the current practice. Future efforts may benefit from combining learning-based modals with classical simulators or statistical models.

\item Open benchmarks are insufficient to compare and reproduce different approaches. Future efforts should focus on public data and standardized platforms.
\end{enumerate}

Even though resilient multi-energy systems have attracted the global attention, opportunities and challenges are intertwined in this emerging area. More research and industrial efforts are needed to address the existing gaps and enable new breakthroughs in securing energy supply under climate change. We hope this review could provide a meaningful step toward in this area.

%

\appendices

\section{Search Queries for Fig.~\ref{fig:bib-stats}} \label{app:search}

We conduct bibliometric analysis on the well-known Web of Science Core Collection database. In the advanced search page, our search query for ``multi-energy system'' and ``multi-energy system resilience'' are given below:

\texttt{ALL = multi energy system OR multi carrier system OR integrated energy system OR energy hub}


\texttt{ALL = (multi energy system OR multi carrier system OR integrated energy system OR energy hub) AND (resilience OR resilient OR extreme weather OR extreme event OR disaster)}

%

Similar search queries were also made on Scopus and Engineering Village, and the major findings still hold.

\section{Full Comparison of the Existing Reviews} \label{app:comp-review}

Our work is unique to investigate climate change impacts and conduct the special interdisciplinary summary of resilient multi-energy systems. This line of research differs from power system resilience~\cite{panteli2015influence,ji2017resilience,huang2024toward}, and it covers more context (climate and financial modeling) than multi-energy system operation~\cite{o2020multicarrier,he2020reliability}. This review formulates a comprehensive framework for multi-energy resilience, which is absent in the related reviews~\cite{jasiunas2021energy,yang2022resilience}.

Table~\ref{tab:comp-review} demonstrates how this paper is different from the relevant reviews with full details. These papers are compared in terms of disturbance sources (cyber attacks, natural disasters, climate change impacts), energy sectors (electricity and other sectors), and modeling details or research tasks (energy hubs, evaluation, operation \& planning, investment, and real-world projects). As shown, the consideration of climate change impacts (climate science), investment decisions (energy finance), and practical implementation separates our work from all the existing reviews. 

\begin{table}
	\caption{Comparison of Different Review Papers}
	\label{tab:comp-review}
	\centering
	\setlength\tabcolsep{3pt}
	\begin{threeparttable} 
		\begin{tabular}{lcccccccc}
			\toprule
			Features              
			& Ours  
			& \cite{panteli2015influence}  
			& \cite{ji2017resilience}
			& \cite{huang2024toward}
			& \cite{o2020multicarrier}
			& \cite{he2020reliability} 
			& \cite{jasiunas2021energy}  
			& \cite{yang2022resilience} \\
			\midrule
			cyber attacks         & /     & /    & /   & yes & /   & /   & yes  & yes \\
			natural disasters     & yes   & yes  & yes & yes & /   & /   & yes  & yes \\
			climate change        & yes   & some & /   & /   & /   & /   & some & /   \\
			\midrule
			electricity sector    & yes   & yes  & yes & yes & yes & yes & yes  & yes \\
			other sectors         & yes   & /    & /   & /   & yes & yes & few  & yes \\
			\midrule
			energy hubs           & yes   & /    & /   & /   & yes & yes & /    & /   \\
			evaluation metrics    & yes   & yes  & yes & /   & /   & yes & yes  & yes \\
			operation \& planning & yes   & yes  & yes & yes & /   & /   & /    & yes \\
			pricing \& investment & yes   & /    & /   & few & /   & /   & /    & /   \\
			real-world projects   & yes   & /    & /   & /   & /   & /   & /    & /  \\
			\bottomrule
		\end{tabular}
	\end{threeparttable}
\end{table}

%


\bibliographystyle{ieeetr}
\bibliography{refs}

\end{document}